\documentclass[prd,showpacs,preprintnumbers,twocolumn,amsmath,nofootinbib,amssymb]{revtex4}
\usepackage{graphicx,color,dcolumn,booktabs,bm}
\usepackage{longtable,lscape}
\usepackage{txfonts}
\usepackage{overpic}
\usepackage{amssymb}
\usepackage{epstopdf}
\usepackage{indentfirst}
\usepackage{feynmf}   
\usepackage{slashed}  
\usepackage{cases}
\usepackage{color}
\usepackage{float}
\usepackage{multirow}
\usepackage{graphicx,color,dcolumn,booktabs,bm}
\usepackage{epsfig,dsfont,amssymb,amsmath,amsfonts,amsbsy,mathrsfs}

\graphicspath{{Figures/}} %
\usepackage[colorlinks, citecolor=blue,anchorcolor=red,menucolor=red, linkcolor=red,filecolor=red,runcolor=red,urlcolor=blue,frenchlinks=red]{hyperref}

\makeatletter
\@addtoreset{equation}{section}
\makeatother

\allowdisplaybreaks
\begin{document}

\title{Exotic triple-charm deuteron-like hexaquarks}

\author{Rui Chen$^{1,2,3}$}
\author{Fu-Lai Wang$^{1,2}$}
\email{chenr15@lzu.edu.cn}
\author{Atsushi Hosaka$^{3}$}
\email{hosaka@rcnp.osaka-u.ac.jp}
\author{Xiang Liu$^{1,2}$}
\email{xiangliu@lzu.edu.cn}
\affiliation{
$^1$School of Physical Science and Technology, Lanzhou University, Lanzhou 730000, China\\
$^2$Research Center for Hadron and CSR Physics, Lanzhou University
and Institute of Modern Physics of CAS, Lanzhou 730000, China\\
$^3$Research Center for Nuclear Physics (RCNP), Osaka University, Ibaraki, Osaka 567-0047, Japan
}

\begin{abstract}
Adopting the one-boson-exchange model, we perform a systematic investigation of interactions between a doubly charmed baryon $(\Xi_{cc})$ and an $S$-wave charmed baryon ($\Lambda_c$, $\Sigma_c^{(*)}$, and $\Xi_c^{(\prime,*)}$). Both the $S$-$D$ mixing effect and coupled-channel effect are considered in this work. Our results suggest that there may exist several possible triple-charm deuteron-like hexaquarks. Meanwhile, we further study the interactions between a doubly charmed baryon and an $S$-wave anticharmed baryon. We find that a doubly charmed baryon and an $S$-wave anticharmed baryon can be easily bound together to form shallow molecular hexaquarks. These heavy flavor hexaquarks predicted here can be accessible at future experiment like LHCb.
\end{abstract}


\maketitle

\section{introduction}\label{sec1}

As a hot issue of hadron physics, exploring exotic hadronic matter is a research field full of challenges and opportunities, which is valuable to deepen  our understanding of non-perturbative behavior of quantum chromodynamics (QCD). The observations of $XYZ$ charmonium-like states, $P_c(4380)$ and $P_c(4450)$ \cite{Aaij:2015tga} have stimulated abundant studies involved in hidden-charm tetraquarks and pentaquarks (see Refs. \cite{Chen:2016qju,Liu:2013waa,Hosaka:2016pey} for review) since 2003.

In general, there are two approaches to clearly identify exotic hadronic states: 1) we may conclude a hadronic state to be an exotic state if it has exotic spin-parity quantum number $J^{PC}$ like $0^{--}$, $0^{+-}$, $1^{-+}$ and so on. A typical candidate is the observation of the $\pi_1(1400)$ from the COMPASS Collaboration \cite{Alde:1988bv} which has $J^{PC}=1^{-+}$, where obviously the $\pi_1(1400)$ cannot be grouped into conventional meson family. 2) If a hadronic state has a typical exotic quark configuration different from the conventional hadron, we may also definitely categorize it as an exotic state. The reported $X(5568)$ with fully open-flavor content $su\bar{c}\bar{d}$ \cite{D0:2016mwd,Abazov:2017poh} is a typical exotic candidate, although no significant is observed by the LHCb collaboration \cite{Aaij:2016iev}, the CMS collaboration of LHC \cite{Sirunyan:2017ofq}, and the CDF collaboration of Fermilab \cite{Aaltonen:2017voc}. The above criteria provides us valuable hints to identify exotic hadronic states.

Very recently, a double-charm baryon $\Xi_{cc}^{++}(3621)$ was discovered by the LHCb Collaboration when analyzing the $\Lambda_c^+K^-\pi^+\pi^-$ invariant mass spectrum \cite{Aaij:2017ueg}. The observed $\Xi_{cc}^{++}(3621)$ not only make the baryon family become complete, but also inspires our interest in exploring the interaction of it with other hadrons, which has a close relation to the exploration of exotic hadronic molecular states.  After the observation of $\Xi_{cc}^{++}(3621)$,
the interaction of two doubly charmed baryons were investigated in Ref. \cite{Meng:2017fwb}, and then the same authors predicted
the possible hadronic molecules composed of the doubly charmed baryon and nucleon \cite{Meng:2017udf}, which is involved in the interaction of a doubly charmed baryon and a nucleon. Recently, Chen, Hosaka, and Liu performed a study of triple-charm molecular pentaquarks by checking the interaction of a doubly charmed baryon and a charmed meson \cite{Chen:2017jjn}. Along this line, it is natural to extend these former studies to the interaction of a doubly charmed baryon and a charmed baryon, which will be a main task of the present work.

\begin{figure}[!htbp]
\center
\includegraphics[width=3.3in]{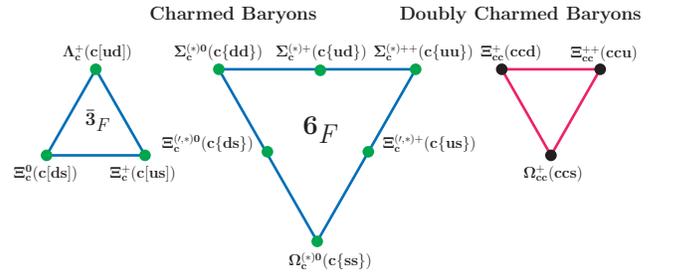}\\
\caption{Charmed baryons in $\bar{3}_F$ and $6_F$ representations and doubly charmed baryons. Here, we define $[q_1q_2]=\frac{1}{\sqrt{2}}(q_1q_2-q_2q_1)$ and $\{q_1q_2\}=\frac{1}{\sqrt{2}}(q_1q_2+q_2q_1)$.}\label{baryon}
\end{figure}

As shown in Fig. \ref{baryon}, charmed baryons can be categorized as $\bar{3}_F$ and $6_F$ representations based on flavor symmetries of light quarks. $\bar{3}_F$ and $6_F$ correspond to the light quarks in flavor antisymmetry and symmetry, respectively. Spin-parity for an $S$-wave charmed baryon is either $1/2^+$ or $3/2^+$. Here, we will discuss the $\Xi_{cc}{\mathcal{B}}(1/2^+)$ interactions with ${\mathcal{B}}=\Lambda_c/\Sigma_c/\Xi_c^{(\prime)}$ and the $\Xi_{cc}{\mathcal{B}}^*(3/2^+)$ interactions with ${\mathcal{B}}^*=\Sigma_c^*/\Xi_{c}^*$.

By examining the interaction of a doubly charmed baryon and a charmed baryon, we want to answer whether or not there exist possible triple-charm deuteron-like hexaquarks, which has the $cccqqq$ configuration. Due to this typical configuration, these discussed triple-charm deuteron-like hexaquarks
are typical exotic hadronic states apparently different from the conventional hadron.
For achieving this goal, in this work we adopt one-boson-exchange (OBE) model to get the effective potential of the interaction of a doubly charmed baryon and a charmed baryon, by which we may search for the corresponding bound state solutions. Finally, we will predict the existence of triple-charm deuteron-like hexaquarks. The present investigation provides crucial information to experimental search for possible triple-charm deuteron-like hexaquarks.

As an extension, exploring the interactions between a doubly charmed baryon and an $S$-wave anticharmed baryon
will be carried out in this work. And then, we further predict possible shallow molecular hexaquarks composed of a doubly charmed baryon and an $S$-wave anticharmed baryon.

This paper is organized as follows. In Sec. \ref{sec2}, the detailed calculation of effective potential related to the interaction between an $S$-wave doubly charmed baryon and an $S$-wave charmed baryon will be given, and the corresponding numerical results will be presented in Sec. \ref{sec3}.
Finally, we will give a short summary in Sec. \ref{sec4}.

\section{Interactions}\label{sec2}

According to the heavy quark symmetry, chiral symmetry and hidden local symmetry \cite{Liu:2011xc,Bando:1984ej}, effective Lagrangians for charmed baryons and light mesons interactions are constructed as
\begin{eqnarray}
\mathcal{L}_{\mathcal{B}_{\bar{3}}} &=& l_B\langle\bar{\mathcal{B}}_{\bar{3}}\sigma\mathcal{B}_{\bar{3}}\rangle
          +i\beta_B\langle\bar{\mathcal{B}}_{\bar{3}}v^{\mu}(\mathcal{V}_{\mu}-\rho_{\mu})\mathcal{B}_{\bar{3}}\rangle,\label{lag1}\\
\mathcal{L}_{\mathcal{B}_{6}} &=&  l_S\langle\bar{\mathcal{S}}_{\mu}\sigma\mathcal{S}^{\mu}\rangle
         -\frac{3}{2}g_1\varepsilon^{\mu\nu\lambda\kappa}v_{\kappa}
         \langle\bar{\mathcal{S}}_{\mu}\mathcal{A}_{\nu}\mathcal{S}_{\lambda}\rangle\nonumber\\
         &&+i\beta_{S}\langle\bar{\mathcal{S}}_{\mu}v_{\alpha}\left(\mathcal{V}^{\alpha}-\rho^{\alpha}\right) \mathcal{S}^{\mu}\rangle
         +\lambda_S\langle\bar{\mathcal{S}}_{\mu}F^{\mu\nu}(\rho)\mathcal{S}_{\nu}\rangle,\label{lag2}\nonumber\\\\
\mathcal{L}_{\mathcal{B}_{\bar{3}}\mathcal{B}_6} &=& ig_4\langle\bar{\mathcal{S}^{\mu}}\mathcal{A}_{\mu}\mathcal{B}_{\bar{3}}\rangle+i\lambda_I\varepsilon^{\mu\nu\lambda\kappa}v_{\mu}\langle \bar{\mathcal{S}}_{\nu}F_{\lambda\kappa}\mathcal{B}_{\bar{3}}\rangle+h.c..\label{lag03}
\end{eqnarray}
Here, $\mathcal{A}_{\mu}$ and $\mathcal{V}_{\mu}$ stand for the axial current and vector current, respectively
\begin{eqnarray*}
\mathcal{A}_{\mu} &=& \frac{1}{2}(\xi^{\dag}\partial_{\mu}\xi-\xi\partial_{\mu}\xi^{\dag})=\frac{i}{f_{\pi}}
\partial_{\mu}{P}+\ldots,\\
\mathcal{V}_{\mu} &=&
\frac{1}{2}(\xi^{\dag}\partial_{\mu}\xi+\xi\partial_{\mu}\xi^{\dag})
=\frac{i}{2f_{\pi}^2}\left[{P},\partial_{\mu}{P}\right]+\ldots
\end{eqnarray*}
with $\xi=\text{exp}(i{P}/f_{\pi})$. $f_{\pi}$ is the pion decay constant with a value of 132 MeV. $\rho^{\mu}=ig_V{V}^{\mu}/\sqrt{2}$, $F^{\mu\nu}(\rho)=\partial^{\mu}\rho^{\nu}-\partial^{\nu}\rho^{\mu}
+\left[\rho^{\mu},\rho^{\nu}\right]$. In Eqs. (\ref{lag2}) and (\ref{lag03}), we define a superfield $\mathcal{S}$, which is expressed as a combination of $\mathcal{B}_6$ with $J^P=1/2^+$ and $\mathcal{B}^*_6$ with $J^P=3/2^+$, $\mathcal{S}_{\mu} =-\sqrt{\frac{1}{3}}(\gamma_{\mu}+v_{\mu})\gamma^5\mathcal{B}_6+\mathcal{B}_{6\mu}^*$. Matrices for $\mathcal{B}_{\bar{3}}$ and $\mathcal{B}_{6}^{(*)}$ are
\begin{eqnarray*}
\mathcal{B}_{\bar{3}} = \left(\begin{array}{ccc}
        0    &\Lambda_c^+      &\Xi_c^+\\
        -\Lambda_c^+       &0      &\Xi_c^0\\
        -\Xi_c^+      &-\Xi_c^0     &0
\end{array}\right),
\mathcal{B}_6^{(*)} = \left(\begin{array}{ccc}
         \Sigma_c^{{(*)}++}                  &\frac{\Sigma_c^{{(*)}+}}{\sqrt{2}}     &\frac{\Xi_c^{(',*)+}}{\sqrt{2}}\\
         \frac{\Sigma_c^{{(*)}+}}{\sqrt{2}}      &\Sigma_c^{{(*)}0}    &\frac{\Xi_c^{(',*)0}}{\sqrt{2}}\\
         \frac{\Xi_c^{(',*)+}}{\sqrt{2}}    &\frac{\Xi_c^{(',*)0}}{\sqrt{2}}      &\Omega_c^{(*)0}
\end{array}\right).
\end{eqnarray*}
In addition, matrices for pseudoscalar mesons ${P}$ and vector mesons ${V}$ are written as
\begin{eqnarray*}
P &=& \left(\begin{array}{ccc}
\frac{\pi^0}{\sqrt{2}}+\frac{\eta}{\sqrt{6}} &\pi^+ &K^+ \nonumber\\
\pi^- &-\frac{\pi^0}{\sqrt{2}}+\frac{\eta}{\sqrt{6}} &K^0 \nonumber\\
K^- &\bar{K}^0 &-\sqrt{\frac{2}{3}}\eta
\end{array}\right),\nonumber\\
V &=& \left(\begin{array}{ccc}
\frac{\rho^0}{\sqrt{2}}+\frac{\omega'}{\sqrt{6}} &\rho^+ &K^{*+} \nonumber\\
\rho^- &-\frac{\rho^0}{\sqrt{2}}+\frac{\omega'}{\sqrt{6}} &K^{*0} \nonumber\\
K^{*-} &\bar{K}^{*0} &-\sqrt{\frac{2}{3}}\omega'
\end{array}\right).
\end{eqnarray*}

\begin{widetext}
By expanded Eqs. (\ref{lag1})$-$(\ref{lag03}), the concrete effective Lagrangians are
\begin{eqnarray}
\mathcal{L}_{\mathcal{B}_{\bar{3}}\mathcal{B}_{\bar{3}}\sigma} &=& l_B\langle \bar{\mathcal{B}}_{\bar{3}}\sigma\mathcal{B}_{\bar{3}}\rangle,\\
\mathcal{L}_{\mathcal{B}_{6}^{(*)}\mathcal{B}_{6}^{(*)}\sigma} &=& -l_S\langle\bar{\mathcal{B}}_6\sigma\mathcal{B}_6\rangle
+l_S\langle\bar{\mathcal{B}}_{6\mu}^{*}\sigma\mathcal{B}_6^{*\mu}\rangle,\\
\mathcal{L}_{\mathcal{B}_{\bar{3}}\mathcal{B}_{\bar{3}}{V}} &=& \frac{1}{\sqrt{2}}\beta_Bg_V\langle\bar{\mathcal{B}}_{\bar{3}}v\cdot{V}\mathcal{B}_{\bar{3}}\rangle,\\
\mathcal{L}_{\mathcal{B}_6^{(*)}\mathcal{B}_6^{(*)}{P}} &=&
        i\frac{g_1}{2f_{\pi}}\varepsilon^{\mu\nu\lambda\kappa}v_{\kappa}\langle\bar{\mathcal{B}}_6
        \gamma_{\mu}\gamma_{\lambda}\partial_{\nu}{P}\mathcal{B}_6\rangle-i\frac{3g_1}{2f_{\pi}}\varepsilon^{\mu\nu\lambda\kappa}v_{\kappa}\langle
\bar{\mathcal{B}}_{6\mu}^{*}\partial_{\nu} {P}\mathcal{B}_{6\lambda}^*\rangle+i\frac{\sqrt{3}}{2}\frac{g_1}{f_{\pi}}v_{\kappa}\varepsilon^{\mu\nu\lambda\kappa}
      \langle\bar{\mathcal{B}}_{6\mu}^*\partial_{\nu}{P}{\gamma_{\lambda}\gamma^5}
      \mathcal{B}_6\rangle+h.c.,\\
\mathcal{L}_{\mathcal{B}_6^{(*)}\mathcal{B}_6^{(*)} {V}} &=& -\frac{\beta_Sg_V}{\sqrt{2}}\langle\bar{\mathcal{B}}_6v\cdot{V}\mathcal{B}_6\rangle-i\frac{\lambda g_V}{3\sqrt{2}}\langle\bar{\mathcal{B}}_6\gamma_{\mu}\gamma_{\nu}
    \left(\partial^{\mu} {V}^{\nu}-\partial^{\nu} {V}^{\mu}\right)
    \mathcal{B}_6\rangle-\frac{\beta_Sg_V}{\sqrt{6}}\langle\bar{\mathcal{B}}_{6\mu}^*v\cdot {V}\left(\gamma^{\mu}+v^{\mu}\right)\gamma^5\mathcal{B}_6\rangle\nonumber\\
    &&-i\frac{\lambda_Sg_V}{\sqrt{6}}\langle\bar{\mathcal{B}}_{6\mu}^*
    \left(\partial^{\mu} {V}^{\nu}-\partial^{\nu} {V}^{\mu}\right)
    \left(\gamma_{\nu}+v_{\nu}\right)\gamma^5\mathcal{B}_6\rangle+\frac{\beta_Sg_V}{\sqrt{2}}\langle\bar{\mathcal{B}}_{6\mu}^*v\cdot {V}\mathcal{B}_6^{*\mu}\rangle+i\frac{\lambda_Sg_V}{\sqrt{2}}\langle\bar{\mathcal{B}}_{6\mu}^*
    \left(\partial^{\mu} {V}^{\nu}-\partial^{\nu} {V}^{\mu}\right)
    \mathcal{B}_{6\nu}^*\rangle+h.c.,\\
\mathcal{L}_{\mathcal{B}_{\bar{3}}\mathcal{B}_6^{(*)}{V}} &=&
       -\frac{\lambda_Ig_V}{\sqrt{6}}\varepsilon^{\mu\nu\lambda\kappa}v_{\mu}\langle \bar{\mathcal{B}}_6\gamma^5\gamma_{\nu}
        \left(\partial_{\lambda} {V}_{\kappa}-\partial_{\kappa} {V}_{\lambda}\right)\mathcal{B}_{\bar{3}}\rangle-\frac{\lambda_Ig_V}{\sqrt{2}}\varepsilon^{\mu\nu\lambda\kappa}v_{\mu}\langle \bar{\mathcal{B}}_{6\nu}^*
          \left(\partial_{\lambda}{V}_{\kappa}-\partial_{\kappa}{V}_{\lambda}\right)
          \mathcal{B}_{\bar{3}}\rangle+h.c.,\\
\mathcal{L}_{\mathcal{B}_{\bar{3}}\mathcal{B}_6^{(*)} {P}} &=& -\sqrt{\frac{1}{3}}\frac{g_4}{f_{\pi}}\langle\bar{\mathcal{B}}_6\gamma^5
\left(\gamma^{\mu}+v^{\mu}\right)\partial_{\mu}{P}\mathcal{B}_{\bar{3}}\rangle-\frac{g_4}{f_{\pi}}\langle\bar{\mathcal{B}}_{6\mu}^*\partial^{\mu} {P}\mathcal{B}_{\bar{3}}\rangle+h.c..
\end{eqnarray}
\end{widetext}

Suppose the interaction between heavy and light quarks is negligible, effective Lagrangians for the $S$-wave doubly charmed baryons and light mesons interactions can be constructed as
\begin{eqnarray}
\mathcal{L}_{\Xi_{cc}\Xi_{cc}\sigma} &=& g_{\sigma}\bar{\Xi}_{cc}\sigma\Xi_{cc},\label{lag3}\\
\mathcal{L}_{\Xi_{cc}\Xi_{cc}{P}} &=& g_{\pi}\bar{\Xi}_{cc}i\gamma_5{P}\Xi_{cc},\\
\mathcal{L}_{\Xi_{cc}\Xi_{cc}{V}} &=& h_{v}\bar{\Xi}_{cc}\gamma_{\mu}{V}^{\mu}\Xi_{cc}
          +\frac{f_v}{2M_{\Xi_{cc}}}\bar{\Xi}_{cc}\sigma_{\mu\nu}\partial^{\mu}{V}^{\nu}\Xi_{cc}.\label{lag4}
          \nonumber\\
\end{eqnarray}

To obtain consistent coupling constants in these two kinds of effective Lagrangians, we would like to borrow the experience from the nucleon-nucleon interaction, which has the form of
\begin{eqnarray}
\mathcal{L}_N &=& g_{\sigma NN}\bar{N}\sigma N+\sqrt{2}g_{\pi NN} \bar{N}i\gamma_5{P}N\nonumber\\
                 &&+\sqrt{2}g_{\rho NN}\bar{N}\gamma_{\mu}{V}^{\mu}N+\frac{f_{\rho NN}}{\sqrt{2}m_N}\bar{N}\sigma_{\mu\nu}\partial^{\mu}{V}^{\nu}N.\label{lag5}
\end{eqnarray}
All of the coupling constants for the charmed baryon, doubly charmed baryon, and nucleon sectors are related in the quark level, the detailed derivations were given in Refs. \cite{Meng:2017fwb,Liu:2011xc}.
In Table \ref{coupling}, we finally summarize the values of all of the coupling constants and hadron masses adopted in the following calculations.

\renewcommand\tabcolsep{0.6cm}
\renewcommand{\arraystretch}{1.8}
\begin{table*}[!htbp]
  \caption{A summary of coupling constants and hadron masses adopted in our calculations. Here, the values relevant to the nucleon-nucleon interactions are given in Refs. \cite{Machleidt:2000ge,Machleidt:1987hj,Cao:2010km}. Masses of the hadrons involved in our study are taken from the Particle Data Group \cite{Olive:2016xmw}. Unit of hadrons masses is MeV.}\label{coupling}
  \begin{tabular}{llll}\toprule[1pt]
  $\frac{g_{\sigma NN}^2}{4\pi}=5.69$      &$\frac{g_{\pi NN}^2}{4\pi}=13.60$       &$\frac{g_{\rho NN}^2}{4\pi}=0.84$     &$\frac{f_{\rho NN}}{g_{\rho NN}}=6.10$\\
  $l_S=-2l_B=-\frac{2}{3}g_{\sigma NN}$    &$g_1=\frac{2\sqrt{2}}{3}g_4=-\frac{2\sqrt{2}f_{\pi}g_{\pi NN}}{5 M_N}$         &$\beta_Sg_V=-2\beta_Bg_V=-4g_{\rho NN}$   &$\lambda_Sg_V=-\sqrt{8}\lambda_Ig_V=-\frac{6(g_{\rho NN}+f_{\rho NN})}{5M_N}$\\
  $g_{\sigma}=\frac{1}{3}g_{\sigma NN}$    &$g_{\pi}=-\frac{\sqrt{2}m_{\Xi_{cc}}g_{\pi NN}}{5 m_N}$     &$h_{v}=\sqrt{2}g_{\rho NN}$      &$h_{v}+f_{v}=-\frac{\sqrt{2}}{5}\frac{m_{\Xi_{cc}}}{m_N}(g_{\rho NN}+f_{\rho NN})$\\\hline
  $m_{\sigma}=600$       &$m_{\pi}=137.27$    &$m_{\eta}=547.85$       &$m_{\rho}=775.49$\\
  $m_{\omega}=782.65$    &$m_N=938.27$        &$m_{\Xi_{cc}}=3621.4$   &$m_{\Lambda_c}=2286.46$\\
  $m_{\Xi_c}=2469.34$     &$m_{\Sigma_c}=2453.54$       &$m_{\Xi_c^{\prime}}=2576.75$                  &$m_{\Sigma_c^*}=2518.07$,\quad $m_{\Xi_c^*}=2645.9$\\
  \bottomrule[1pt]
  \end{tabular}
\end{table*}

Under a Breit approximation, the effective potentials in momentum space is related to the corresponding scattering amplitude, i.e.,
\begin{eqnarray}
\mathcal{V}_{E}^{h_1h_2\to h_3h_4}(\bm{q}) &=&
          -\frac{\mathcal{M}(h_1h_2\to h_3h_4)}
          {\sqrt{\prod_i2M_i\prod_f2M_f}}.
\end{eqnarray}
$\mathcal{M}(h_1h_2\to h_3h_4)$, $M_i$, and $M_f$ correspond to the scattering amplitude for the $h_1h_2\to h_3h_4$ process, the masses of the initial states ($h_1$, $h_2$) and final states ($h_3$, $h_4$), respectively.
For the effective potentials in coordinate space $\mathcal{V}(r)$, it is obtained by performing a Fourier transformation,
\begin{eqnarray}
\mathcal{V}_{E}^{h_1h_2\to h_3h_4}({r}) =
          \int\frac{d^3\bm{q}}{(2\pi)^3}e^{i\bm{q}\cdot\bm{r}}
          \mathcal{V}_{E}^{h_1h_2\to h_3h_4}(\bm{q})\mathcal{F}^2(q^2,m_E^2).\nonumber\\
\end{eqnarray}
Because the discussed hadrons are not point-like particles, here, we introduce a monopole form factor\footnote{In general, the other kinds of form factor (like dipole form factor and exponential type form factor) are also adopted to discuss the hadron-hadron interaction. Once expanded these form factors in the powers of $q^2/\Lambda^2$, $\mathcal{F}(q^2)\sim 1+c_1\times q^2/\Lambda^2+c_2\times(q^2/\Lambda^2)^2+\ldots$, different form factors correspond to different sets of coefficients. Nevertheless, in the low momentum limit, all of these different coefficients can be absorbed into the interactive strengthes. Since the physics of molecular states is essentially low momentum physics, numerical results are quite similar by suitably choosing cutoffs and coupling constants.} $\mathcal{F}(q^2,m_E^2)=(\Lambda^2-m_E^2)/(\Lambda^2-q^2)$ at each interaction vertexes, which is often adopted to study the  nucleon-nucleon interaction \cite{Tornqvist:1993ng,Tornqvist:1993vu}. $\Lambda$, $m_E$, and $q$ stand for cutoff, mass, and four-momentum of exchanged mesons, respectively. In general, cutoff $\Lambda$ is related to the typical hadronic scale or the intrinsic size of hadrons. In our former paper \cite{Chen:2017jjn}, we reproduced the bound state property for the deuteron with the same parameters, by employing the cutoff $\Lambda = 0.862$ GeV.
Assuming that the intrinsic size of hadrons are similar to that of the nucleon, we employ in the present study the cutoff $\Lambda$ of the same order around 1 GeV.

Here, we need to emphasis that all of the interaction strengthes determined from the nucleon-nucleon interaction, and the only one phenomenological parameter, cutoff $\Lambda$, is estimated from the deuteron. In the following calculation, we will vary cutoff $\Lambda$ from 0.8 GeV to 5.0 GeV to search for the loosely bound solutions. The bound state with its cutoff close to 1 GeV may be the possible molecular candidate.

To deduce total effective potentials, we need further construct wave functions for all of the discussed systems, which include the spin-orbit wave function, the flavor wave function, and the spatial wave function. In Table \ref{flavor}, the flavor wave functions are collected.

\renewcommand\tabcolsep{0.42cm}
\renewcommand{\arraystretch}{1.7}
\begin{table}[!htbp]
  \caption{Flavor wave functions for the discussed triple-charm hexaquark systems composed of a S-wave double-charm baryon and an $S$-wave charmed baryon.}\label{flavor}
  \begin{tabular}{lll}\toprule[1pt]
  Systems     &$|I,I_3\rangle$     &Configurations\\\midrule[1pt]
  $\Xi_{cc}\Lambda_c$
          &$\left|\frac{1}{2},\frac{1}{2}\right\rangle$    &$\left|\Xi_{cc}^{++}\Lambda_c^+\right\rangle$\\

          &$\left|\frac{1}{2},-\frac{1}{2}\right\rangle$    &$\left|\Xi_{cc}^{+}\Lambda_c^+\right\rangle$
\\\hline
  $\Xi_{cc}\Xi_{c}^{(\prime,*)}$
          &$\left|1,1\right\rangle$      &$\left|\Xi_{cc}^{++}\Xi_c^{(\prime,*)+}\right\rangle$\\
          &$\left|1,0\right\rangle$
              &$\frac{1}{\sqrt{2}}\left(\left|\Xi_{cc}^{++}\Xi_c^{(\prime,*)0}\right\rangle
                      +\left|\Xi_{cc}^+\Xi_{c}^{(\prime,*)+}\right\rangle\right)$\\
          &$\left|1,-1\right\rangle$   &$\left|\Xi_{cc}^{+}\Xi_c^{(\prime,*)0}\right\rangle$\\
          &$\left|0,0\right\rangle$
              &$\frac{1}{\sqrt{2}}\left(\left|\Xi_{cc}^{++}\Xi_c^{(\prime,*)0}\right\rangle
                      -\left|\Xi_{cc}^+\Xi_{c}^{(\prime,*)+}\right\rangle\right)$
          \\\hline
    $\Xi_{cc}\Sigma_c^{(*)}$
          &$\left|\frac{1}{2},\frac{1}{2}\right\rangle$
               &$\frac{1}{\sqrt{3}}\left|\Xi_{cc}^{++}\Sigma_c^{(*)+}\right\rangle
                 -\sqrt{\frac{2}{3}}\left|\Xi_{cc}^+\Sigma_c^{(*)++}\right\rangle$\\
          &$\left|\frac{1}{2},-\frac{1}{2}\right\rangle$
                &$\sqrt{\frac{2}{3}}\left|\Xi_{cc}^{++}\Sigma_c^{(*)0}\right\rangle
                  -\frac{1}{\sqrt{3}}\left|\Xi_{cc}^{+}\Sigma_c^{(*)+}\right\rangle$\\
          &$\left|\frac{3}{2},\frac{3}{2}\right\rangle$
                &$\left|\Xi_{cc}^{++}\Sigma_c^{(*)++}\right\rangle$\\
          &$\left|\frac{3}{2},\frac{1}{2}\right\rangle$
                &$\sqrt{\frac{2}{3}}\left|\Xi_{cc}^{++}\Sigma_c^{+(*)}\right\rangle
                       +\frac{1}{\sqrt{3}}\left|\Xi_{cc}^{+}\Sigma_c^{(*)++}\right\rangle$\\
          &$\left|\frac{3}{2},-\frac{1}{2}\right\rangle$
                &$\frac{1}{\sqrt{3}}\left|\Xi_{cc}^{++}\Sigma_c^{(*)0}\right\rangle
                       +\sqrt{\frac{2}{3}}\left|\Xi_{cc}^{+}\Sigma_c^{(*)+}\right\rangle$\\
          &$\left|\frac{3}{2},-\frac{3}{2}\right\rangle$
                &$\left|\Xi_{cc}^+\Sigma_c^{(*)0}\right\rangle$\\

  \bottomrule[1pt]
  \end{tabular}
\end{table}

Since the $S$-$D$ mixing effect is considered in this work, their spin-orbit wave functions include
\begin{eqnarray*}
\Xi_{cc}\mathcal{B}:  &J^P=0^+&  \left|{}^1S_0\right\rangle,\\
                           &J^P=1^+&  \left|{}^3S_1\right\rangle,\quad   \left|{}^3D_1\right\rangle,\\
\Xi_{cc}\mathcal{B}^*:  &J^P=1^+&  \left|{}^3S_1\right\rangle,\quad   \left|{}^3D_1\right\rangle,\quad
                                      \left|{}^5D_1\right\rangle,\\
                           &J^P=2^+&  \left|{}^5S_2\right\rangle,\quad   \left|{}^3D_2\right\rangle,\quad
                                      \left|{}^5D_2\right\rangle
\end{eqnarray*}
with
\begin{eqnarray*}
\Xi_{cc}\mathcal{B} \left|{}^{2S+1}L_{J}\right\rangle &=& \sum_{m,n,m_L}C_{\frac{1}{2},m;\frac{1}{2},n}^{S,ms}C_{S,m_S;L,m_L}^{J,m_J}
       \chi_{\frac{1}{2},m}\chi_{\frac{1}{2},n}Y_{L,m_L},\\
\Xi_{cc}\mathcal{B}^* \left|{}^{2S+1}L_{J}\right\rangle &=&
\sum_{m,n,m_L}C_{\frac{1}{2},m;\frac{3}{2},n}^{S,m_S}C_{S,m_S;L,m_L}^{J,m_J}
       \chi_{\frac{1}{2},m}\Phi_{\frac{3}{2},n}Y_{L,m_L}.
\end{eqnarray*}
Here, $C_{\frac{1}{2},m;\frac{1}{2},n}^{S,ms}$, $C_{\frac{1}{2},m;\frac{3}{2},n}^{S,m_S}$, and $C_{S,m_S;L,m_L}^{J,m_J}$ are the Clebsch-Gordan coefficients. $\chi_{\frac{1}{2},m}$ and $\Phi_{\frac{3}{2},n}$ are defined as the spin wave functions for baryons with spin $1/2$ and spin $3/2$, respectively. And $\Phi_{\frac{3}{2},n}=\sum_{n_1,n_2}\langle\frac{1}{2},n_1;1,n_2|\frac{3}{2},n\rangle
\chi_{\frac{1}{2},n_1}\epsilon^{n_2}$. The polarization vector $\epsilon$ has the form of $\epsilon_{\pm}^{m}=\mp\frac{1}{\sqrt{2}}\left(\epsilon_x^{m}{\pm}i\epsilon_y^{m}\right)$ and $\epsilon_0^{m}=\epsilon_z^{m}$ with $\epsilon_{\pm1}= \frac{1}{\sqrt{2}}\left(0,\pm1,i,0\right)$ and $\epsilon_{0} =\left(0,0,0,-1\right)$.

\section{Numerical results}\label{sec3}

After getting all of the effective potentials for these discussed systems composed of a $S$-wave doubly charmed baryon and an $S$-wave charmed baryon (see  Appendix \ref{app01} for more details), we solve the coupled channel Schr\"{o}dinger equation and try to find the corresponding bound state solutions (binding energy $E$ and root-mean-square (RMS) radius $r_{RMS}$).

In the following, we present the results for single- and coupled-channel cases, separately.

\subsection{Single-channel Case}\label{single}

\subsubsection{$\Xi_{cc}\Lambda_c$ and $\Xi_{cc}\Xi_c$ systems}\label{seca1}

\begin{figure}[!htbp]
\center
\includegraphics[width=3.4in]{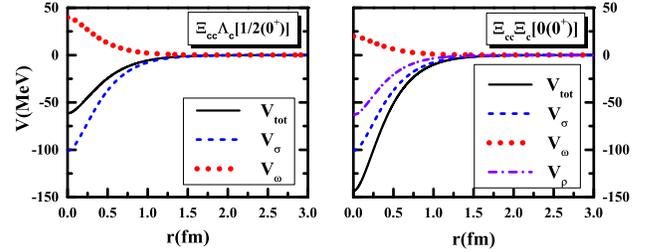}
\caption{The $r$ dependence of the deduced effective potentials for the $\Xi_{cc}\Lambda_c$ system with $I(J^P)=1/2(0^+)$ and the $\Xi_{cc}\Xi_c$ system with $I(J^P)=0(0^+)$. Here,  we take the cutoff $\Lambda=1.00$ GeV}\label{lam}
\end{figure}

The forbidden $\pi/\eta-\Lambda_c-\Lambda_c$ and $\pi/\eta-\Xi_c-\Xi_c$ interactions explain why there don't exist the $\pi/\eta$-exchange potentials for the $\Xi_{cc}\Lambda_c$ and $\Xi_{cc}\Xi_c$ systems. Additionally, the $\rho$-exchange potential is also absent for the $\Xi_{cc}\Lambda_c$ system since it is forbidden by the isospin symmetry. The properties for the OBE effective potentials
from other remaining exchanged mesons have been constructed in Refs. \cite{Chen:2017vai,Chen:2017xat}. The interactions from the $\sigma$, $\omega$, and $\rho$ exchanges depend on the quark configurations and the isospin of the concrete hadron-hadron systems. Here, the quark configuration for the $\Xi_{cc}\Lambda_c$ and $\Xi_{cc}\Xi_c$ systems is $(ccq)$-$(cqq)$ with $q=u,d,s$. Thus, the $\sigma$ exchange always provides an attractive force, and the  $\omega$ exchange has contribution to repulsive potential. Because the $\rho$ couples to the isospin charge, the effective potential from the $\rho$ exchange is strongly attractive for the isoscalar $\Xi_{cc}\Xi_c$ system, and weakly repulsive for the isovector $\Xi_{cc}\Xi_c$ system. In Fig. \ref{lam}, we present the $r$ dependence of the effective potentials for the $\Xi_{cc}\Lambda_c$ system with $I(J^P)=1/2(0^+)$ and the $\Xi_{cc}\Xi_c$ system with $I(J^P)=0(0^+)$, where the cutoff $\Lambda$ is taken as 1 GeV.

\begin{figure}[!htbp]
\center
\includegraphics[width=3.4in]{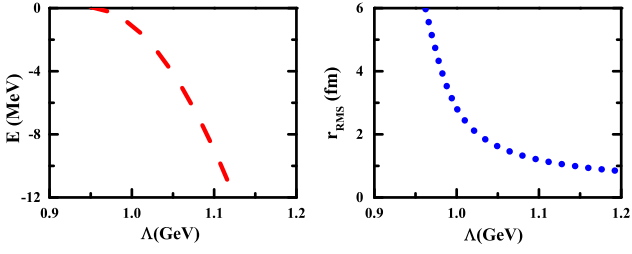}
\caption{Bound state solutions (binding energy $E$ and RMS radius $r_{RMS}$) for the $\Xi_{cc}\Xi_c$ system with $I(J^P)=0(0^+/1^+)$.}\label{xi}
\end{figure}

By tuning cutoff value $\Lambda$ from 0.8 GeV to 5 GeV, we cannot find bound solutions for the $\Xi_{cc}\Lambda_c$ systems and the isovector $\Xi_{cc}\Xi_c$ systems. In Fig. \ref{xi}, we present the corresponding bound state properties for the isoscalar $\Xi_{cc}\Xi_c$ systems.Here, when cutoff $\Lambda$ is taken around 1 GeV, binding energy of several MeV is obtained, and the RMS radius is larger than 1 fm, which is consistent with a typical size of a hadron-hadron molecular state. Thus, there is the possibility that the isoscalar $\Xi_{cc}\Xi_c$ states with $J^P=0^+, 1^+$ can be good candidates of triple-charm molecular hexaquarks, depending on the actual value of the cutoff $\Lambda$.

The suggested decay channels of the predicted isoscalar $\Xi_{cc}\Xi_c$ states with $J^P=0^+, 1^+$ are $\Omega_{cc}\Lambda_c$ and $\Omega_{ccc}\Lambda$. Due to the absence of $\Omega_{ccc}$ in experiment, there still exists big challenge for searching for these predicted triple-charm molecular hexaquarks.

\subsubsection{$\Xi_{cc}\Sigma_c^{(*)}$ and $\Xi_{cc}\Xi_{c}^{(\prime,*)}$ systems}

For the $\Xi_{cc}\Sigma_c^{(*)}$ and $\Xi_{cc}\Xi_{c}^{(\prime,*)}$ systems, the $\pi$ exchange is allowed, which plays important role to the $\Xi_{cc}\Sigma_c^{(*)}$ and $\Xi_{cc}\Xi_{c}^{(\prime,*)}$ molecular systems.

\begin{figure}[h
!]
\center
\includegraphics[width=3.4in]{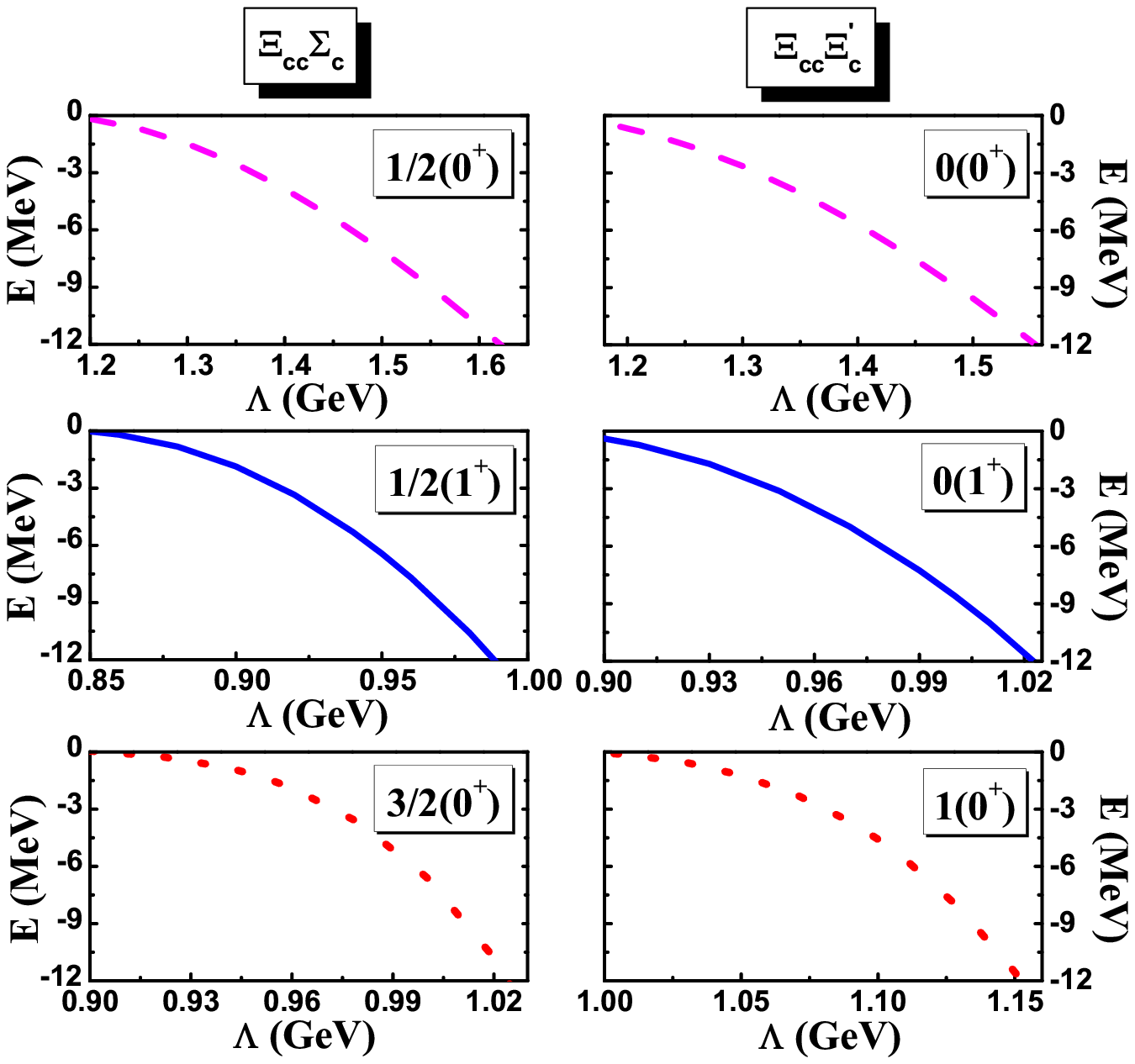}
\includegraphics[width=3.4in]{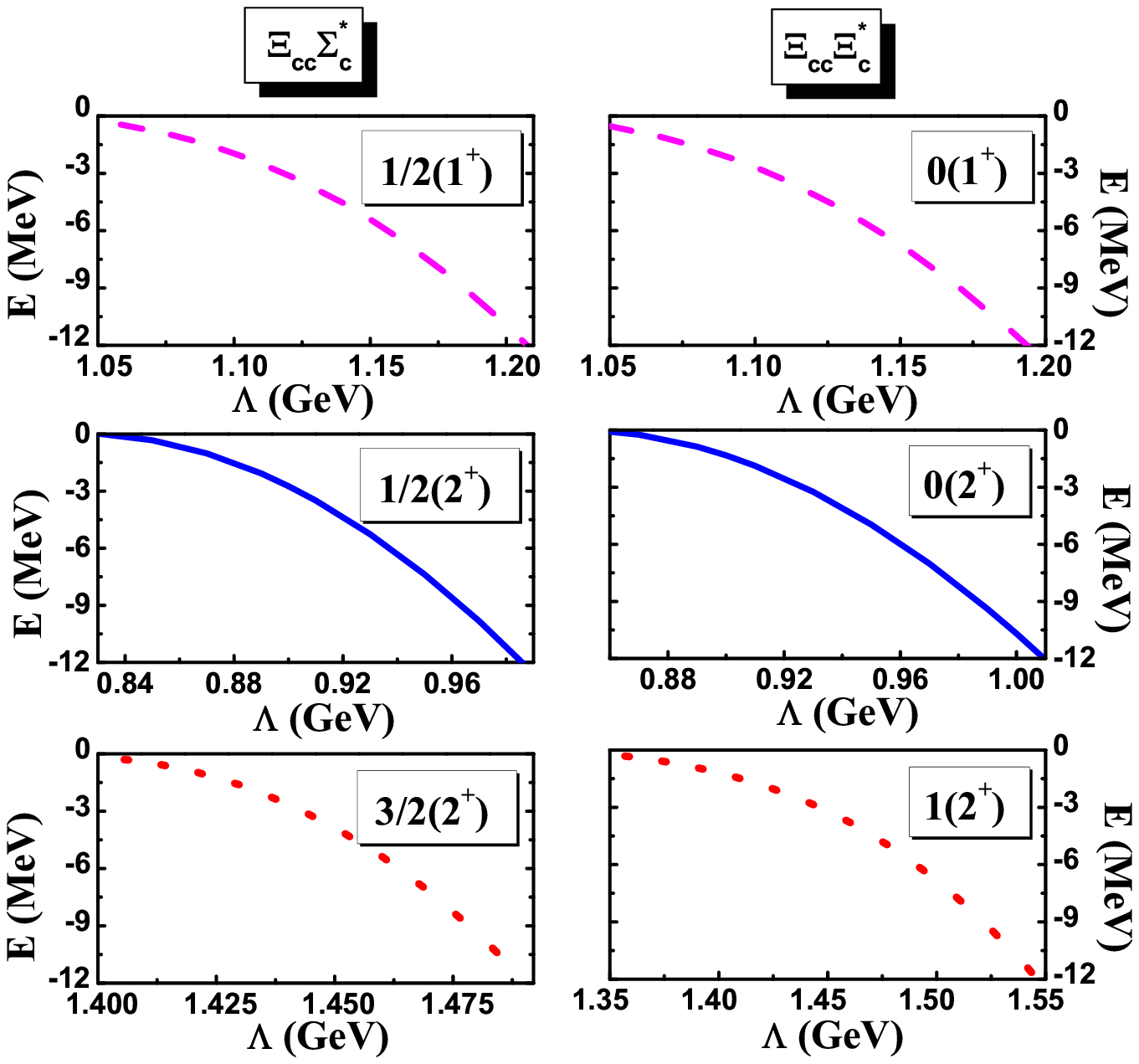}
\caption{$\Lambda$ dependence of binding energy $E$ for the $\Xi_{cc}\Sigma_c^{(*)}$ and $\Xi_{cc}\Xi_{c}^{(\prime,*)}$ systems.}\label{b6}
\end{figure}

In Fig. \ref{b6}, we present the $\Lambda$ dependence of the binding energies of these $\Xi_{cc}\Sigma_c^{(*)}$ and $\Xi_{cc}\Xi_{c}^{(\prime,*)}$ systems, which shows that
there exist bound state solutions for the $\Xi_{cc}\Sigma_c^{(*)}$ and $\Xi_{cc}\Xi_{c}^{(\prime,*)}$ systems with all allowed isospin and spin-parity quantum numbers.
Our results suggest that several systems may be possible candidates of triple-charm molecular hexaquarks, which are the $\Xi_{cc}\Sigma_c$ states with $I(J^P)=1/2(0^+, 1^+)$, $3/2(0^+)$, the $\Xi_{cc}\Sigma_c^*$ states with $I(J^P)=1/2(2^+, 1^+)$, $3/2(2^+)$, the $\Xi_{cc}\Xi_c^{\prime}$ states with $I(J^P)=0(0^+, 1^+)$, $1(0^+)$, and the $\Xi_{cc}\Xi_c^*$ states with $I(J^P)=0(1^+, 2^+)$, $1(2^+)$.

For the $\Xi_{cc}\Sigma_c^{(*)}$ systems with $I(J^P)=3/2(1^+)$ and the $\Xi_{cc}\Xi_{c}^{(\prime,*)}$ systems with $I(J^P)=1(1^+)$, when cutoff is tuned from 2 to 3 GeV, we can obtain a binding energy around
a few to 10 MeV.

For providing more abundant information of experimental search for them, in the following we further discuss their decay behaviors. Possible two-body strong decay channels are $\Xi_{cc}\Lambda_c$, $\Xi_{cc}\Sigma_c$, $\Omega_{ccc}N$, and $\Omega_{ccc}\Delta$ for the predicted $\Xi_{cc}\Sigma_c^{(*)}$ molecular states. For the $\Xi_{cc}\Xi_{c}^{(\prime,*)}$ molecular states, their allowed decay modes include the $\Xi_{cc}\Xi_c^{(')}$, $\Omega_{ccc}\Lambda$, and $\Omega_{ccc}\Sigma$ channels.

\subsection{Coupled-channel Case}\label{couple}

For the study of coupled channels, necessary channels are summarized in Table \ref{channel}. For the ispspin $I=3/2$ case, there are only two systems ($\Xi_{cc}\Sigma_c$ and $\Xi_{cc}\Sigma_c^*$) to be considered here.

\renewcommand\tabcolsep{0.1cm}
\renewcommand{\arraystretch}{1.7}
\begin{table}[!htbp]
\caption{Possible channels involved in the coupled-channel investigation.}\label{channel}
{\begin{tabular}{c|lll}
\toprule[1pt]
$I(J^P)$      &\multicolumn{3}{c}{Channels}\\\hline
$1/2(0^+)$    &$\Xi_{cc}\Lambda_c\left|{}^1S_0\right\rangle$                  &$\Xi_{cc}\Sigma_c\left|{}^1S_0\right\rangle$                &$\Xi_{cc}\Sigma_c^*\left|{}^5D_0\right\rangle$\\
$1/2(1^+)$    &$\Xi_{cc}\Lambda_c\left|{}^3S_1, {}^3D_1\right\rangle$         &$\Xi_{cc}\Sigma_c\left|{}^3S_1, {}^3D_1\right\rangle$       &$\Xi_{cc}\Sigma_c^*\left|{}^3S_1, {}^3D_1, {}^5D_1\right\rangle$\\
$3/2(0^+)$    &\ldots     &$\Xi_{cc}\Sigma_c\left|{}^1S_0\right\rangle$                   &$\Xi_{cc}\Sigma_c^*\left|{}^5D_0\right\rangle$\\
$3/2(1^+)$    &\ldots     &$\Xi_{cc}\Sigma_c\left|{}^3S_1, {}^3D_1\right\rangle$       &$\Xi_{cc}\Sigma_c^*\left|{}^3S_1, {}^3D_1, {}^5D_1\right\rangle$\\
$0,1(0^+)$    &$\Xi_{cc}\Xi_c\left|{}^1S_0\right\rangle$                  &$\Xi_{cc}\Xi_c^{\prime}\left|{}^1S_0\right\rangle$                &$\Xi_{cc}\Xi_c^*\left|{}^5D_0\right\rangle$\\
$0,1(1^+)$    &$\Xi_{cc}\Xi_c\left|{}^3S_1, {}^3D_1\right\rangle$         &$\Xi_{cc}\Xi_c^{\prime}\left|{}^3S_1, {}^3D_1\right\rangle$       &$\Xi_{cc}\Xi_c^*\left|{}^3S_1, {}^3D_1, {}^5D_1\right\rangle$\\
\bottomrule[1pt]
\end{tabular}}
\end{table}

\begin{figure}[!htbp]
\center
\includegraphics[width=3.4in]{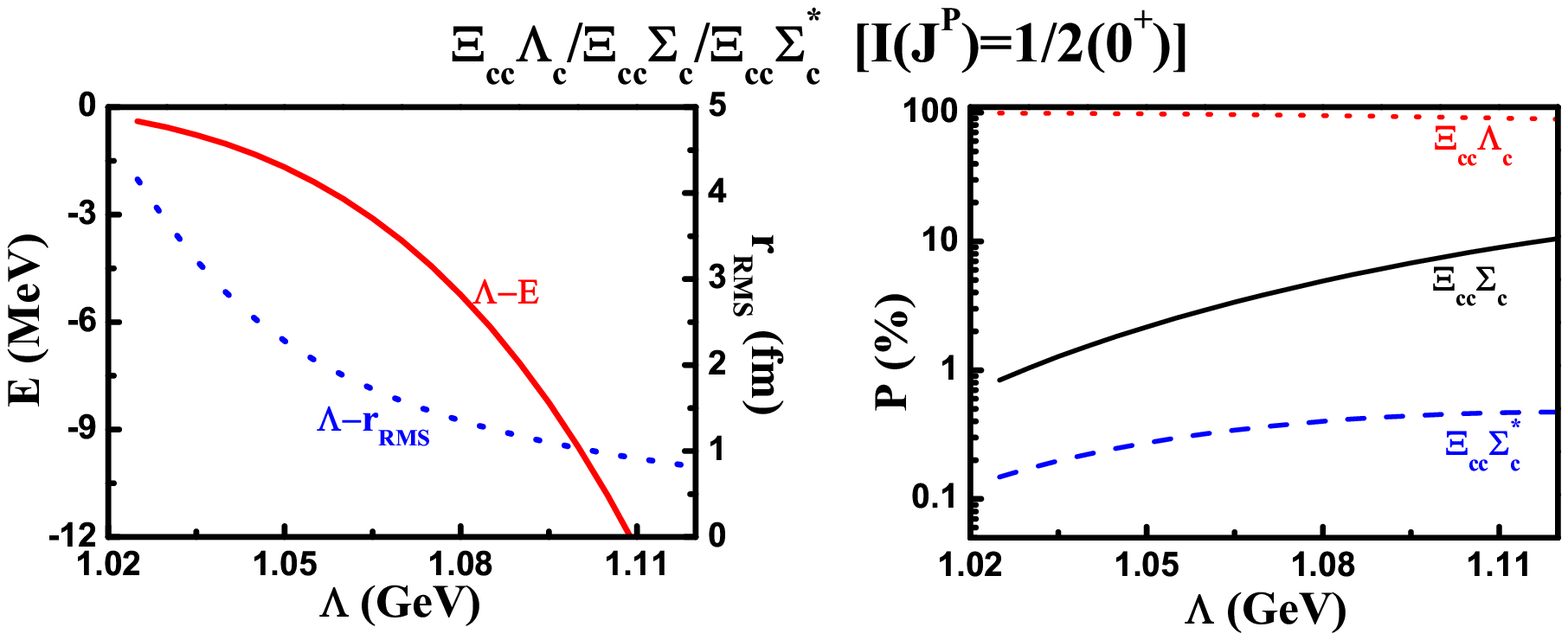}
\includegraphics[width=3.4in]{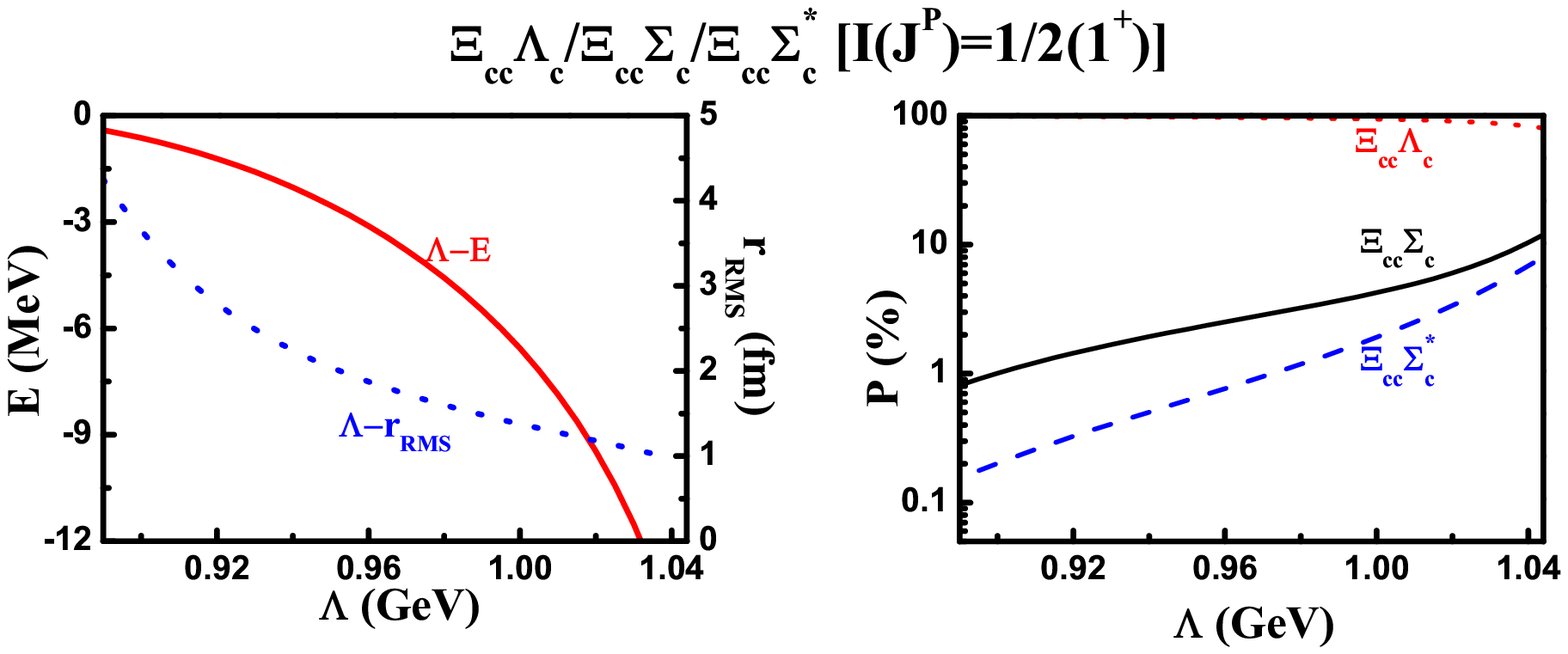}
\includegraphics[width=3.4in]{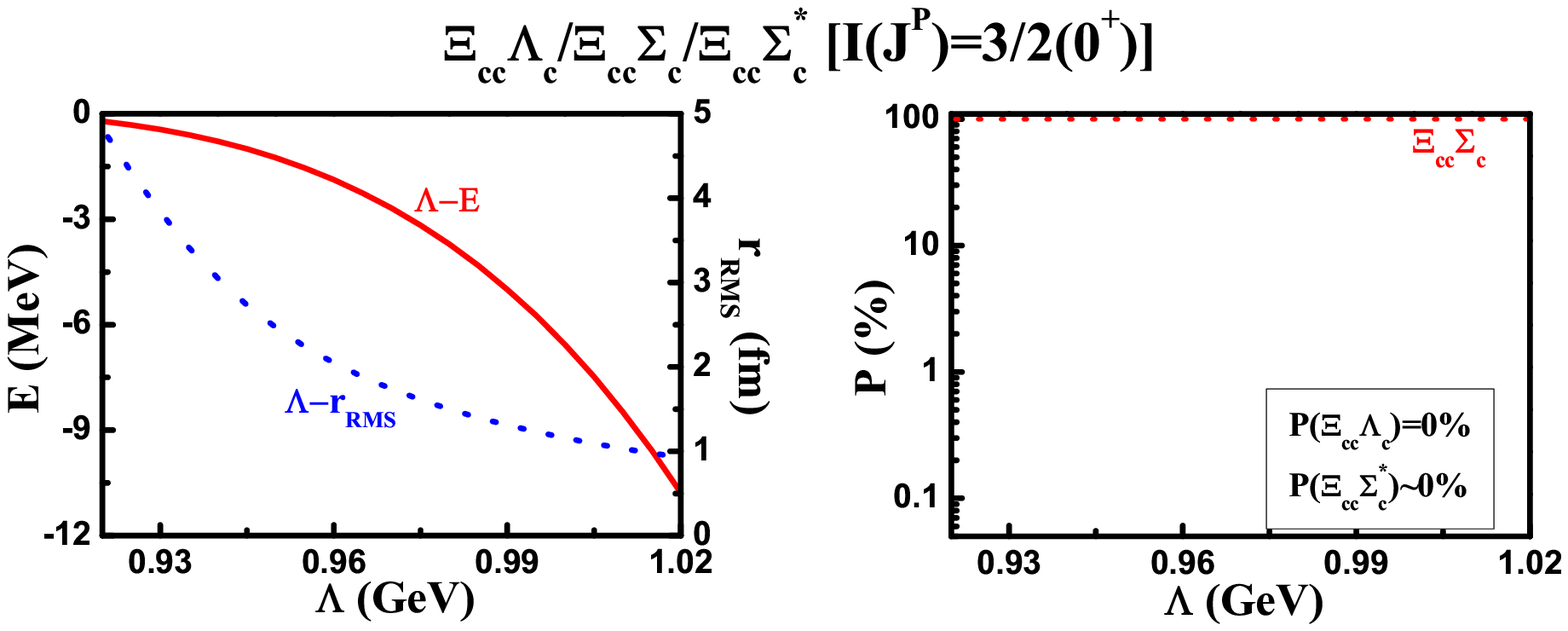}
\includegraphics[width=3.4in]{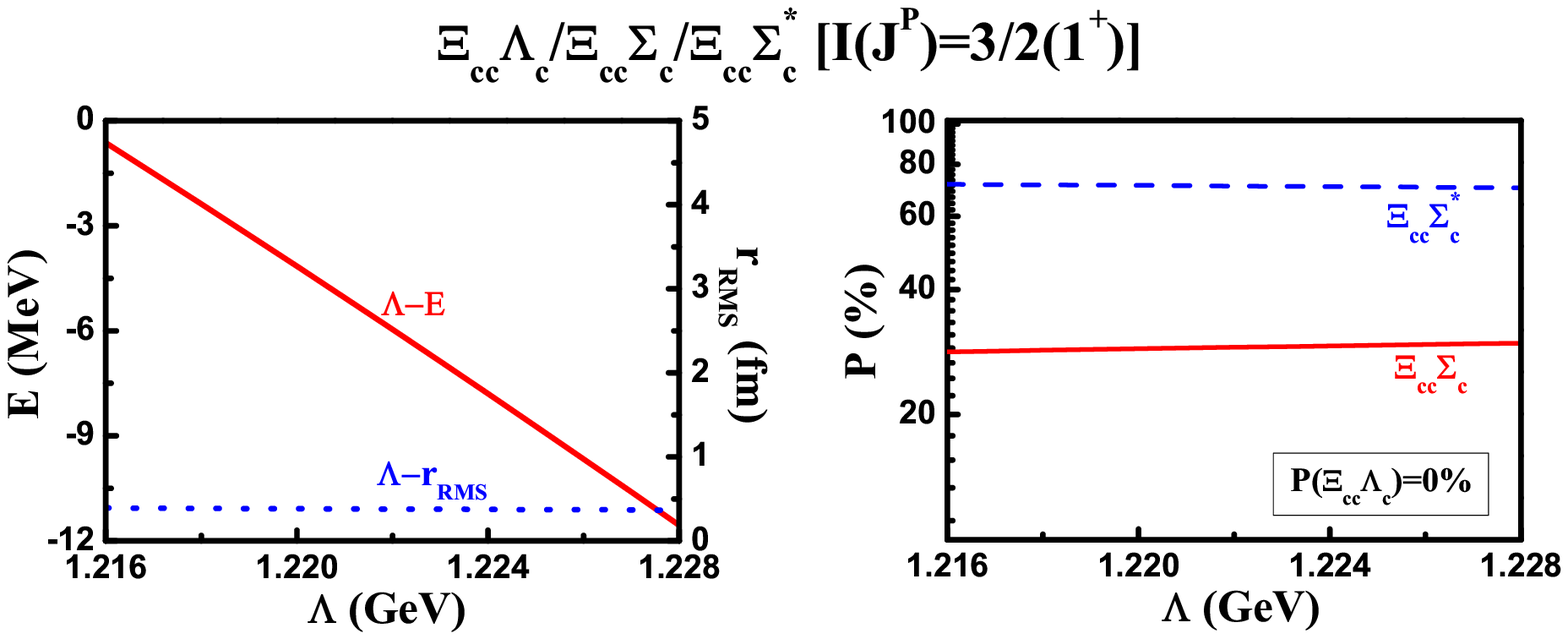}
\caption{Left: Bound state properties (binding energy $E$ and root-mean-square radius $r_{RMS}$) for all of the $\Xi_{cc}\Lambda_c/\Xi_{cc}\Sigma_c/\Xi_{cc}\Sigma_c^*$ systems when the coupled channel effect is included in our calculation. Right: Probability for the different channels.}\label{couplelam}
\end{figure}

In Fig. \ref{couplelam}, we present the bound state solutions for the investigated $\Xi_{cc}\Lambda_c/\Xi_{cc}\Sigma_c/\Xi_{cc}\Sigma_c^*$ systems when considering the coupled-channel effect, where the cutoff $\Lambda$ is taken to be around 1 GeV.  The results for the $\Xi_{cc}\Lambda_c/\Xi_{cc}\Sigma_c/\Xi_{cc}\Sigma_c^*$ coupled systems with $I(J^P)=$ $1/2(0^+)$, $1/2(1^+)$, $3/2(0^+)$, and $3/2(1^+)$ quantum numbers are obtained.

\begin{figure}[!htbp]
\center
\includegraphics[width=3.4in]{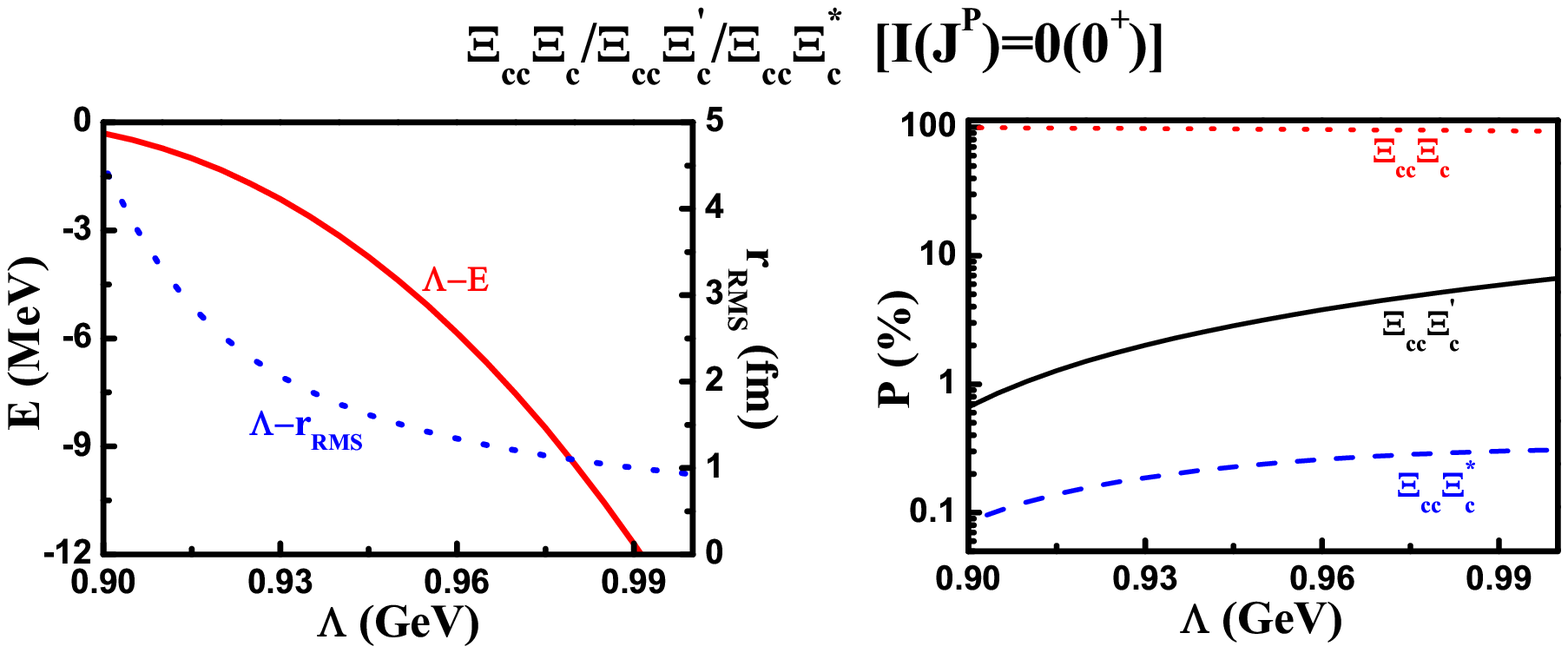}
\includegraphics[width=3.4in]{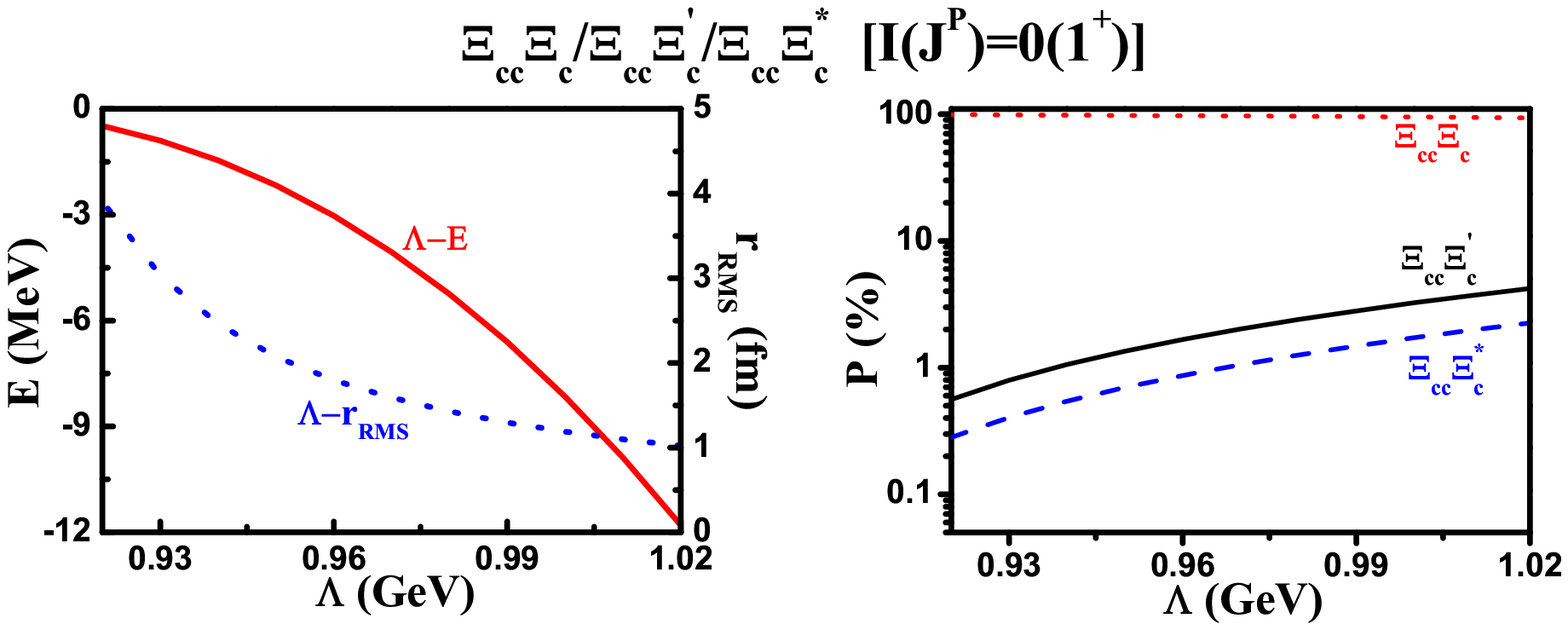}
\includegraphics[width=3.4in]{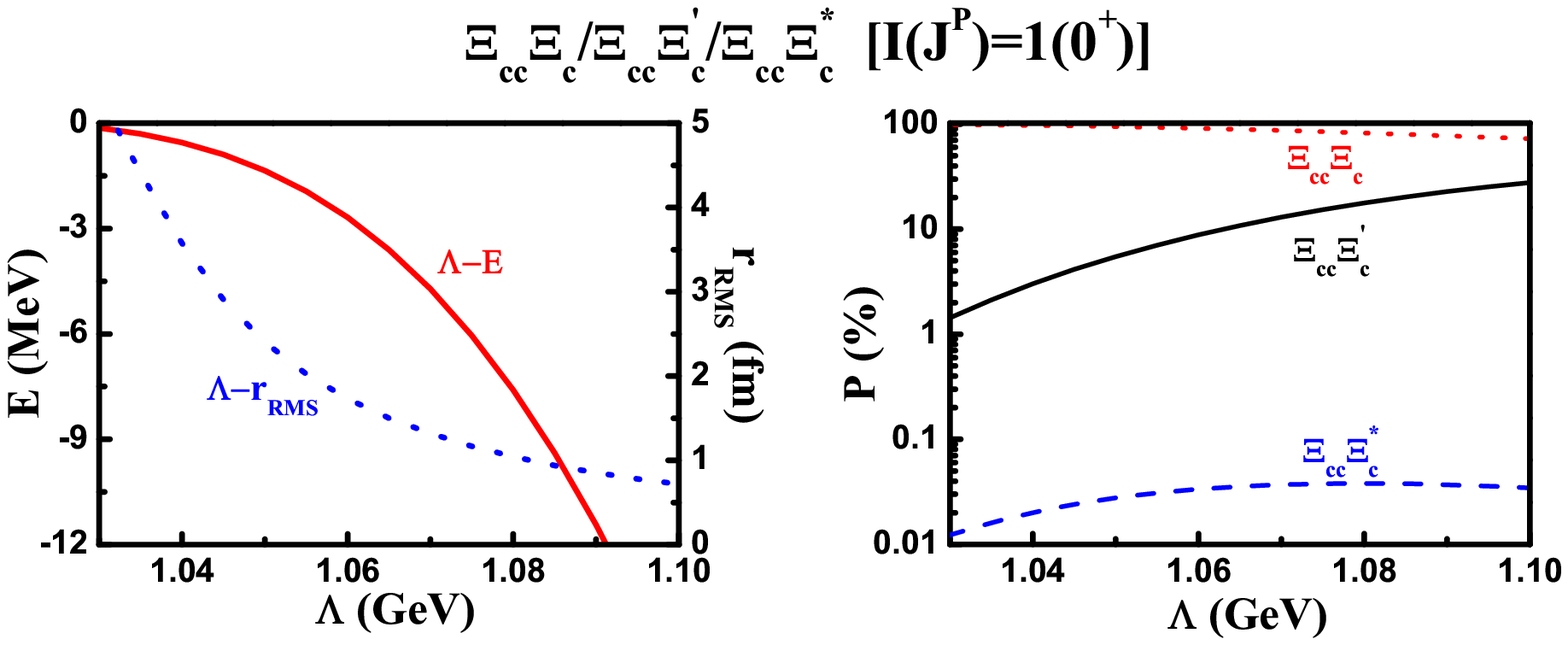}
\includegraphics[width=3.4in]{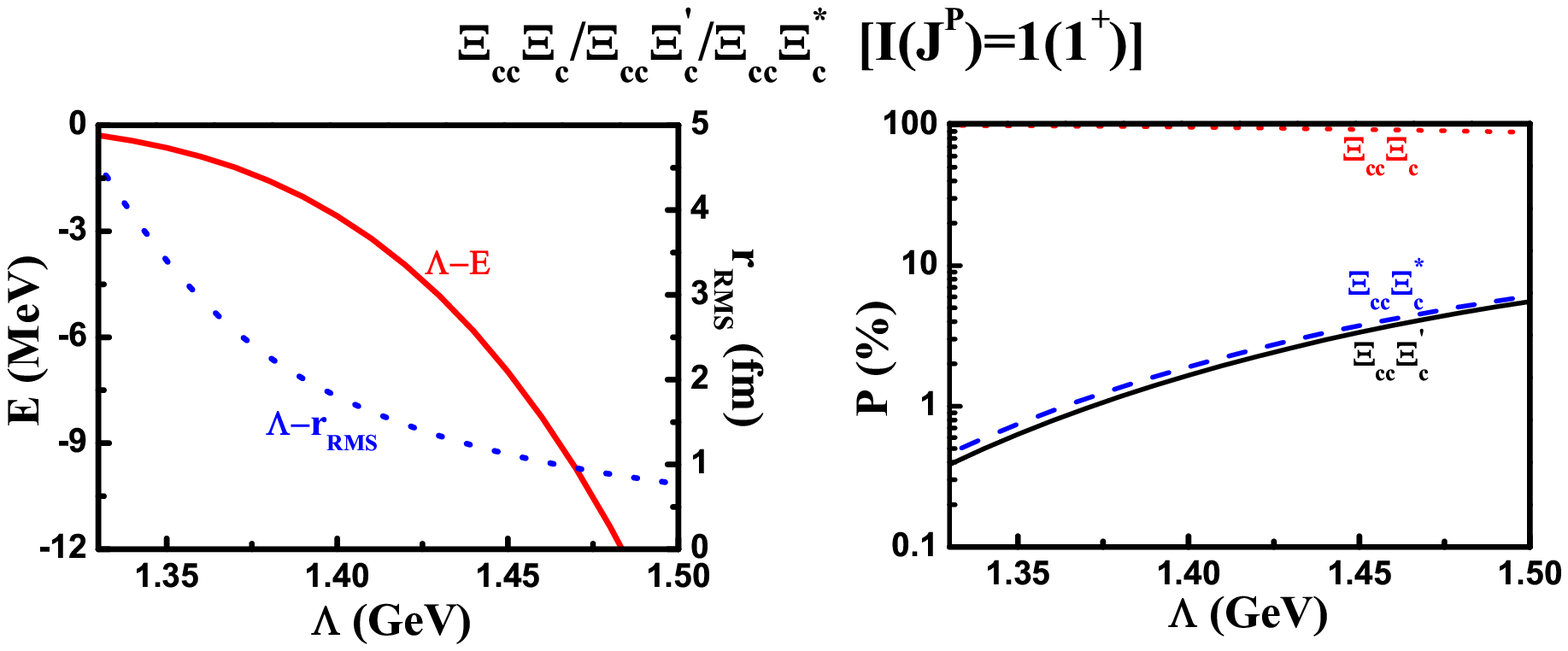}
\caption{Left: Bound state properties (binding energy $E$ and root-mean-square radius $r_{RMS}$) for all of the $\Xi_{cc}\Xi_c/\Xi_{cc}\Xi_c^{\prime}/\Xi_{cc}\Xi_c^*$ systems when the coupled channel effect is included in our calculation. Right: Probability for the different channels.}\label{coupleXi}
\end{figure}

{For the $\Xi_{cc}\Lambda_c/\Xi_{cc}\Sigma_c/\Xi_{cc}\Sigma_c^*$ coupled system with $I(J^P)=$ $1/2(0^+, 1^+)$, according to the selected cutoff value ($\Lambda\sim 1$ GeV) and the obtained RMS radius ($r_{RMS}>1$ fm), we find that the present meson-exchange approach support both of them may be the loosely bound molecular candidates. Because the dominant channel is the $\Xi_{cc}\Lambda_c$ system with probabilities over 90\%, they are mainly composed of the $\Xi_{cc}\Lambda_c$ system.} Compared to the results for the single $\Xi_{cc}\Lambda_c$ system shown in Sec. \ref{seca1}, here, we find that the coupled-channel effect plays a very important role to generate these two loosely molecular states. However, for the $\Xi_{cc}\Lambda_c/\Xi_{cc}\Sigma_c/\Xi_{cc}\Sigma_c^*$ coupled system with $I(J^P)=$ $3/2(0^+)$, our study shows that the coupled-channel effect is not very important. In fact, the bound state solutions do not change very much from the results for the single $\Xi_{cc}\Sigma_c$ channel with $I(J^P) = 3/2(0^+)$, and the probability for the $\Xi_{cc}\Sigma_c$ channel is almost 100\% as presented in Fig. \ref{couplelam}.

{For the $\Xi_{cc}\Lambda_c/\Xi_{cc}\Sigma_c/\Xi_{cc}\Sigma_c^*$ system with $I(J^P)=3/2(1^+)$ shown in Fig. \ref{couplelam}, our result indicates that its RMS radius is as small as 0.37 fm, and the dominant channel is the $\Xi_{cc}\Sigma_c^*$ channel with a probability around 70\%. It seems that the $\Xi_{cc}\Lambda_c/\Xi_{cc}\Sigma_c/\Xi_{cc}\Sigma_c^*$ system with $I(J^P)=3/2(1^+)$ cannot be a reasonable loosely bound state but a deeply bound state.}

In the following, we discuss the $\Xi_{cc}\Xi_c/\Xi_{cc}\Xi_c^{\prime}/\Xi_{cc}\Xi_c^*$ coupled-channel systems. For the $I(J^P)=0(0^+,1^+)$ case, we also find that the contribution of the coupled-channel effect is not obvious since the obtained property of
this coupled-channel system is similar to that of the single $\Xi_{cc}\Xi_c$ states with $I(J^P)=0(0^+,1^+)$. However, for the isovector $\Xi_{cc}\Xi_c/\Xi_{cc}\Xi_c^{\prime}/\Xi_{cc}\Xi_c^*$  systems, after considering the coupled-channel effect, their bound state properties are obtained, which means that there may exist
isovector $\Xi_{cc}\Xi_c/\Xi_{cc}\Xi_c^{\prime}/\Xi_{cc}\Xi_c^*$ molecular states with $J^P=0^+$ and $1^+$.

To summarize, after considering the coupled-channel effect, we may predict the existence of four molecular hexaquarks, which are the $\Xi_{cc}\Lambda_c/\Xi_{cc}\Sigma_c/\Xi_{cc}\Sigma_c^*$ coupled-channel system with $I(J^P)=1/2(0^+, 1^+)$, where the $\Xi_{cc}\Lambda_c$ channel is dominant,
and  the $\Xi_{cc}\Xi_c/\Xi_{cc}\Xi_c^{\prime}/\Xi_{cc}\Xi_c^*$ coupled systems with $I(J^P)=1(0^+,1^+)$, which mainly couple with the $\Xi_{cc}\Xi_c$ channel.

\subsection{Predictions of other molecular hexaquarks}\label{partner}

In this section, we will extend our study to the interaction between a doubly charmed baryon and an $S$-wave anticharmed baryon. Their OBE effective potentials can be related to the effective potentials for the $\Xi_{cc}{\mathcal{B}}^{(*)}$ systems by a $G$-parity rule \cite{Klempt:2002ap}, i.e.,
\begin{eqnarray}
\mathcal{V}^{A\bar{B}\to A\bar{B}} &=& \sum_E {G_E}\mathcal{V}_E^{A{B}\to A{B}},
\end{eqnarray}
where $G_E$ stands for the $G$-parity for the exchanged meson $E$.

\renewcommand\tabcolsep{0.15cm}
\renewcommand{\arraystretch}{1.7}
\begin{table*}[!htbp]
\caption{Bound state properties (binding energy $E$ and root-mean-square radius $r_{RMS}$) for the molecular hexaquarks composed of a doubly charmed baryon and an $S$-wave anticharmed baryon. Here, $E$, $r_{RMS}$, and $\Lambda$ are in units of MeV, fm, and GeV, respectively.}\label{num1}
{\begin{tabular}{cccc|cccc||cccc|cccc}
\toprule[1pt]
$I(J^P)$                &$\Lambda$    &$E$           &$r_{RMS}$        &$I(J^P)$      &$\Lambda$    &$E$     &$r_{RMS}$
&$I(J^P)$                &$\Lambda$    &$E$           &$r_{RMS}$        &$I(J^P)$      &$\Lambda$    &$E$     &$r_{RMS}$\\\hline
\multicolumn{8}{c||}{$\Xi_{cc}\bar{\Lambda}_c$}           &\multicolumn{8}{c}{$\Xi_{cc}\bar{\Xi}_c$}\\\hline
$1/2(0^+/1^+)$  &1.00         &$-0.48$         &4.00             &&&&&$0(0^+/1^+)$  &0.95         &$-0.78$         &3.26              &$1(0^+/1^+)$  &1.10         &$-0.39$         &4.20\\
                        &1.10         &$-6.77$         &1.35             &&&&&              &1.00         &$-5.09$         &1.50              &              &1.30         &$-6.20$         &1.36\\
                        &1.20         &$-19.44$        &0.90             &&&&&              &1.05         &$-13.06$        &1.04              &              &1.50         &$-16.50$        &0.92\\\midrule[1pt]
\multicolumn{4}{c|}{$\Xi_{cc}\bar{\Sigma}_c$}         &\multicolumn{4}{c||}{$\Xi_{cc}\bar{\Sigma}^*_c$}
            &\multicolumn{4}{c|}{$\Xi_{cc}\bar{\Xi}^{\prime}_c$}                   &\multicolumn{4}{c}{$\Xi_{cc}\bar{\Xi}^{*}_c$}\\\hline
$1/2(0^+)$      &0.80         &$-0.84$         &3.01            &$1/2(1^+)$      &0.80         &-$0.41$         &4.01
          &$0(0^+)$      &0.85         &$-0.95$         &2.87           &$0(1^+)$      &0.85         &$-1.08$         &2.72\\
                        &0.95         &$-7.46$         &1.27            &&0.84         &$-3.86$         &1.60
          &              &0.95         &$-8.72$         &1.20           &              &0.90         &$-7.14$         &1.24\\
                        &1.10         &$-10.97$        &1.18            &&0.88         &$-12.52$        &1.01
          &              &1.05         &$-19.52$        &0.93           &              &0.95         &$-19.73$        &0.84\\\hline
$1/2(1^+)$      &0.92         &$-0.57$         &3.76            &$1/2(2^+)$      &0.95         &$-0.14$         &5.60
        &$0(1^+)$      &0.95         &$-0.78$         &3.27             &$0(2^+)$      &1.00         &$-0.64$         &3.60\\
                        &0.96         &$-4.74$         &1.59            &&1.05         &$-4.93$         &1.65
        &              &1.00         &$-5.78$         &1.44             &              &1.50         &$-6.90$         &1.31\\
                        &1.00         &$-13.93$        &1.06            &&1.15         &$-15.19$        &1.12
        &              &1.05         &$-16.19$        &0.98             &              &1.80         &$-16.64$        &0.94\\\hline
$3/2(0^+)$      &1.35         &$-0.20$         &5.14            &$3/2(1^+)$      &1.00         &$-0.59$         &3.57
             &$1(0^+)$      &1.20         &$-0.14$         &5.43     &$1(1^+)$      &1.20         &$-0.49$         &3.89\\
                        &1.70         &$-4.14$         &1.64            &&1.55         &$-4.40$         &1.60
             &              &1.50         &$-4.71$         &1.54     &              &1.50         &$-4.71$         &1.54\\
                        &2.05         &$-9.94$         &1.16            &&1.85         &$-11.84$        &1.08
             &              &1.80         &$-11.84$        &1.07     &              &1.80         &$-11.84$        &1.07\\\hline
$3/2(1^+)$      &1.00         &$-0.59$         &3.57               &$3/2(2^+)$      &1.00         &$-0.71$         &3.32
          &$1(1^+)$      &1.10         &$-1.75$         &2.26          &$1(2^+)$      &1.00         &$-0.18$         &5.09\\
                        &1.10         &$-4.31$         &1.56               &&1.10         &$-4.42$         &1.54
          &              &1.20         &$-5.84$         &1.37          &              &1.15         &$-5.16$         &1.43\\
                        &1.20         &$-11.00$        &1.08               &&1.20         &$-10.73$        &1.08
          &              &1.30         &$-11.81$        &1.04          &              &1.30         &$-15.42$        &0.93\\
\bottomrule[1pt]
\end{tabular}}
\end{table*}
With these obtained effective potentials, we get the corresponding results as presented in Table \ref{num1}.

If setting the cutoff to be around 1 GeV, we can find bound state solutions for all of the discussed systems composed of a doubly charmed baryon and an $S$-wave anticharmed baryon. We find their RMS radius are around 1 fm, which is a typical size of the molecular state with small binding energy.
These predicted molecular states have typical quark configuration $cc\bar{c}\bar{q}\bar{q}q$. Here, we also list the possible allowed decay channels, i.e.,
$DD\bar{D}$, $J/\psi D\pi(\eta)$, $D_sD\bar{D}$, $J/\psi D_s\pi(\eta)$, $J/\psi DK$, which provide valuable information to further experimental exploration to them.

\section{summary}\label{sec4}

Exploring the exotic state is an interesting research topic. Especially with the experimental progress on charmoniumum-like $XYZ$ states and $P_c(4380)/P_c(4450)$ states, theorists have paid more attentions to
the study of exotic states like hidden-charm molecular states and compact multiquarks (see review papers \cite{Chen:2016qju,Liu:2013waa} for more details).

In 2017, a doubly charmed baryon $\Xi_{cc}^{++}(3621)$ was discovered by LHCb \cite{Aaij:2017ueg}.
This new observation makes the study of the interaction between a doubly charmed baryon and an $S$-wave charmed baryon become possible. In this work, we focus on this typical exotic hadronic configuration. By applying the OBE model, we extract their effective potentials, by which we try to find their bound state solutions by solving the Schr\"odinger equation. This information is crucial to conclude whether there exist the corresponding triple-charm molecular hexaquarks. In the present study, the $S$-$D$ mixing effect  and the coupled-channel effect are taken into account. In this work, all of the coupling constants are determined from the nucleon-nucleon interaction and by using the valence quark structure of charmed baryons. In this regard, the molecular hadrons of doubly charmed hadrons allow to study the interactions of a single valence quark. Cutoff $\Lambda$ is roughly estimated around 1 GeV, which is also widely accepted as a reasonable input from the experience of studying the deuteron in Refs. \cite{Tornqvist:1993ng,Tornqvist:1993vu}.

To clarify the uncertainty of cutoff $\Lambda$, we present the $\Lambda$ dependence of the bound properties for all the possible molecular candidates in last section. Obviously, it cannot be molecular candidates as binding energies depend very sensitively on the cutoff parameter, like the coupled $\Xi_{cc}\Lambda_c/\Xi_{cc}\Sigma_c/\Xi_{cc}\Sigma_c^*$ system with $I(J^P)=3/2(1^+)$.

Finally, our discussions are at the qualitative level, we should admit that we cannot make very quantitative predictions. Nevertheless, we imply a serial of possible triple-charm molecular hexaquark states, which are summarized in Table \ref{num2}. When making comparison of the results with and without the coupled-channel effect,
we find that the coupled-channel effect plays an essential role for some discussed coupled-channel systems.

\renewcommand\tabcolsep{0.08cm}
\renewcommand{\arraystretch}{1.7}
\begin{table}[!htbp]
\caption{A summary of predicted hexaquark molecular states.}\label{num2}
{\begin{tabular}{ll|ll}
\toprule[1pt]
States     &$I(J^P)$     &States     &$I(J^P)$\\\hline
$\Xi_{cc}\Lambda_c$       &$1/2(0^+)$,  $1/2(1^+)$                             &$\Xi_{cc}\Xi_c$               &$0(0^+)$, $0(1^+)$, $1(0^+)$, $1(1^+)$\\
$\Xi_{cc}\Sigma_c$        &$1/2(0^+)$, $1/2(1^+)$, $3/2(0^+)$                  &$\Xi_{cc}\Xi_c^{\prime}$      &$0(0^+)$, $0(1^+)$, $1(0^+)$\\
$\Xi_{cc}\Sigma_c^*$      &$1/2(1^+)$, $1/2(2^+)$, $3/2(2^+)$                  &$\Xi_{cc}\Xi_c^*$             &$0(1^+)$, $0(2^+)$, $1(2^+)$\\
\bottomrule[1pt]
\end{tabular}}
\end{table}

As a byproduct, we further extend our study to the interaction between a doubly charmed baryon and an $S$-wave anticharmed baryon since its effective potential can be related to that of the system composed of
a doubly charmed baryon and an $S$-wave charmed baryon by $G$-parity rule. Furthermore, we could predict the existence of molecular hexaquarks candidates with typical quark configuration $cc\bar{c}\bar{q}\bar{q}q$. The experimental search for them is also an intriguing issue.

About fifteen years ago, the observed $X(3872)$ \cite{Choi:2003ue} stimulated extensive discussion of molecular state constructed by charmed meson pair. Later, in 2015, the observation of two hidden-charm $P_c$ states \cite{Aaij:2015tga} can be suggested to be molecular system compose of  a charmed meson and a charmed baryon. The reported double-charm baryon $\Xi_{cc}^{++}(3621)$ \cite{Aaij:2017ueg} again provides us good chance to study double-charm baryon interacting with other hadrons. Along this line, we carried out a realistic study of triple-charm molecular hexaquarks and predicted the existence of them. In the next decades, we have reason to believe that
the predictions can be accessible at future experiment, which will be full of opportunities and challenges.

\section*{ACKNOWLEDGMENTS}
This project is partly supported by the National Natural Science Foundation of China under Grant No. 11222547, No. 11175073, and the Fundamental Research Funds for the Central Universities. R. C. is supported by the China Scholarship Council. X. L. is also supported in part by the National Program for Support of Top-notch Young Professionals. A. H. is supported by the JSPS KAKENHI [the Grant-in-Aid for Scientific Research from the Japan Society for the Promotion of Science (JSPS)] with Grant No. JP26400273(C).

\appendix*

\section{Relevant subpotentials}\label{app01}

The exact OBE effective potentials for all of the investigated processes are expressed as

\begin{widetext}
\begin{eqnarray}
V^{\Xi_{cc}\Lambda_c\to\Xi_{cc}\Lambda_c}_{I=1/2} &=& -2g_{\sigma}l_BY(\Lambda,m_{\sigma},r)+\frac{1}{\sqrt{2}}h_v\beta_Bg_V Y(\Lambda,m_{\omega},r),\\
\mathcal{V}_{I=1/2}^{\Xi_{cc}\Lambda_c\to\Xi_{cc}\Sigma_c} &=& -\frac{\sqrt{2}}{12}\frac{g_{\pi}g_1}{f_{\pi}m_{\Xi_{cc}}}\left[\mathcal{A}_1\mathcal{O}_r
+\mathcal{A}_2\mathcal{P}_r\right] Y(\Lambda_0,m_{\pi0},r)
-\frac{\lambda_Ig_V(h_V-f_V)}{6m_{\Xi_{cc}}}\left[2\mathcal{A}_1\mathcal{O}_r
-\mathcal{A}_2\mathcal{P}_r\right]
Y(\Lambda_0,m_{\rho0},r)\nonumber\\
&&+\frac{\lambda_Ig_Vh_V}{m_{\Xi_{cc}}}\mathcal{A}_{3}\mathcal{Q}_r
Y(\Lambda_0,m_{\rho0},r),\\
\mathcal{V}_{I=1/2}^{\Xi_{cc}\Lambda_c\to\Xi_{cc}\Sigma_c^*} &=& -\frac{1}{2\sqrt{6}}\frac{g_{\pi}g_4}{f_{\pi}m_{\Xi_{cc}}}\left[\mathcal{A}_{8}\mathcal{O}_r
+\mathcal{A}_{9}\mathcal{P}_r\right] Y(\Lambda_1,m_{\pi1},r)
+\frac{\sqrt{3}}{2}\frac{\lambda_Ig_Vh_V}{m_{\Xi_{cc}}}
\mathcal{A}_{10}\mathcal{Q}_rY(\Lambda_1,m_{\rho1},r)\nonumber\\
&&-\frac{\sqrt{3}}{12}\frac{\lambda_Ig_V(h_V-f_V)}{m_{\Xi_{cc}}}
\left[2\mathcal{A}_{8}\mathcal{O}_r
-\mathcal{A}_{9}\mathcal{P}_r\right]Y(\Lambda_1,m_{\rho1},r),\\
V^{\Xi_{cc}\Sigma_c\to\Xi_{cc}\Sigma_c}_I &=& g_{\sigma}l_SY(\Lambda,m_{\sigma},r)
-\frac{\mathcal{H}(I)}{12}\frac{g_{\pi}g_1}{f_{\pi}M_{\Xi_{cc}}}\left(\mathcal{A}_1\mathcal{O}_r
+\mathcal{A}_2\mathcal{P}_r\right)Y(\Lambda,m_{\pi},r)
-\frac{g_{\pi}g_1}{36f_{\pi}M_{\Xi_{cc}}}\left(\mathcal{A}_1\mathcal{O}_r
+\mathcal{A}_2\mathcal{P}_r\right)Y(\Lambda,m_{\eta},r)\nonumber\\
&&-\frac{\mathcal{H}(I)}{2}\left(\frac{ h_v\beta_Sg_V}{\sqrt{2}}
+\frac{h_v\lambda_Sg_V}{3\sqrt{2}M_{\Sigma_c}}\mathcal{O}_r\right)Y(\Lambda,m_{\rho},r)
-\frac{\mathcal{H}(I)}{3\sqrt{2}}h_v\lambda_Sg_V\left(\frac{1}{M_{\Sigma_c}}+\frac{1}{M_{\Xi_{cc}}}\right)
\mathcal{A}_3\mathcal{Q}_rY(\Lambda,m_{\rho},r)\nonumber\\
&&-\frac{\mathcal{H}(I)}{18\sqrt{2}}\frac{(f_v+h_v)\lambda_Sg_V}{M_{\Xi_{cc}}}(2\mathcal{A}_1\mathcal{O}_r
-\mathcal{A}_2\mathcal{P}_r)Y(\Lambda,m_{\rho},r)\nonumber\\
&&-\frac{1}{2}\left(\frac{ h_v\beta_Sg_V}{\sqrt{2}}
+\frac{h_v\lambda_Sg_V}{3\sqrt{2}M_{\Sigma_c}}\mathcal{O}_r\right)Y(\Lambda,m_{\omega},r)
-\frac{1}{3\sqrt{2}}h_v\lambda_Sg_V\left(\frac{1}{M_{\Sigma_c}}+\frac{1}{M_{\Xi_{cc}}}\right)
\mathcal{A}_3\mathcal{Q}_rY(\Lambda,m_{\omega},r)\nonumber\\
&&-\frac{1}{18\sqrt{2}}\frac{(f_v+h_v)\lambda_Sg_V}{M_{\Xi_{cc}}}(2\mathcal{A}_1\mathcal{O}_r
-\mathcal{A}_2\mathcal{P}_r)Y(\Lambda,m_{\omega},r),\\
\mathcal{V}_{I}^{\Xi_{cc}\Sigma_c\to\Xi_{cc}\Sigma_c^*} &=& -\frac{\mathcal{H}(I)}{4\sqrt{3}}\frac{g_{\pi}g_1}{f_{\pi}m_{\Xi_{cc}}}
\left[\mathcal{A}_{11}\mathcal{O}_r+\mathcal{A}_{12}\mathcal{P}_r\right] Y(\Lambda_2,m_{\pi2},r)
-\frac{1}{24\sqrt{3}}\frac{g_{\pi}g_1}{f_{\pi}m_{\Xi_{cc}}}
\left[\mathcal{A}_{11}\mathcal{O}_r+\mathcal{A}_{12}\mathcal{P}_r\right]Y(\Lambda_2,m_{\eta2},r)\nonumber\\
&&+\frac{\mathcal{H}(I)}{\sqrt{6}}\frac{\lambda_Sg_Vh_V}{m_{\Xi_{cc}}}\mathcal{A}_{13}
\mathcal{Q}_rY(\Lambda_2,m_{\rho2},r)
-\frac{\mathcal{H}(I)}{6\sqrt{6}}\frac{\lambda_Sg_V(f_V-h_V)}{m_{\Xi_{cc}}}
\left[2\mathcal{A}_{11}\mathcal{O}_r
-\mathcal{A}_{12}\mathcal{P}_r\right]Y(\Lambda_2,m_{\rho2},r)\nonumber\\
&&+\frac{1}{2\sqrt{6}}\frac{\lambda_Sg_Vh_V}{m_{\Xi_{cc}}}\mathcal{A}_{13}
\mathcal{Q}_rY(\Lambda_2,m_{\omega2},r)
-\frac{1}{12\sqrt{6}}\frac{\lambda_Sg_V(f_V-h_V)}{m_{\Xi_{cc}}}\left[2\mathcal{A}_{11}\mathcal{O}_r
-\mathcal{A}_{12}\mathcal{P}_r\right]Y(\Lambda_2,m_{\omega2},r),\\
V^{\Xi_{cc}\Sigma_c^*\to\Xi_{cc}\Sigma_c^*}_I &=& g_{\sigma}l_S\mathcal{A}_4Y(\Lambda,m_{\sigma},r)
-\frac{\mathcal{H}(I)}{8}\frac{g_{\pi}g_1}{M_{\Xi_{cc}}f_{\pi}}
\left[\mathcal{A}_5\mathcal{O}_r+\mathcal{A}_6\mathcal{P}_r\right]Y(\Lambda,m_{\pi},r)
-\frac{1}{24}\frac{g_{\pi}g_1}{M_{\Xi_{cc}}f_{\pi}}
\left[\mathcal{A}_5\mathcal{O}_r+\mathcal{A}_6\mathcal{P}_r\right]Y(\Lambda,m_{\eta},r)\nonumber\\
&&-\frac{\mathcal{H}(I)}{2\sqrt{2}}h_v\beta_Sg_V
\mathcal{A}_4Y(\Lambda,m_{\rho},r)
-\frac{\mathcal{H}(I)}{2\sqrt{2}}\frac{h_v\lambda_Sg_V}{M_{\Xi_{cc}}}
\mathcal{A}_7\mathcal{Q}_rY(\Lambda,m_{\rho},r)\nonumber\\
&&+\frac{\mathcal{H}(I)}{12\sqrt{2}}\frac{(h_v+f_v)\lambda_sg_V}{M_{\Xi_{cc}}}
\left[2\mathcal{A}_5\mathcal{O}_r-\mathcal{A}_6\mathcal{P}_r\right]
Y(\Lambda,m_{\rho},r)\nonumber\\
&&-\frac{1}{2\sqrt{2}}h_v\beta_Sg_V
\mathcal{A}_4Y(\Lambda,m_{\omega},r)
-\frac{1}{2\sqrt{2}}h_v\lambda_Sg_V
\mathcal{A}_7\mathcal{Q}_rY(\Lambda,m_{\omega},r)\nonumber\\
&&+\frac{1}{12\sqrt{2}}\frac{(h_v+f_v)\lambda_sg_V}{M_{\Xi_{cc}}}
\left[2\mathcal{A}_5\mathcal{O}_r-\mathcal{A}_6\mathcal{P}_r\right]
Y(\Lambda,m_{\omega},r),\\
\mathcal{V}_{I}^{\Xi_{cc}\Xi_c\to\Xi_{cc}\Xi_c} &=& -2g_{\sigma}l_BY(\Lambda,m_{\sigma},r)
+\frac{\mathcal{G}(I)}{2\sqrt{2}}h_v\beta_Bg_V Y(\Lambda,m_{\rho},r)
+\frac{1}{2\sqrt{2}}h_v\beta_Bg_V Y(\Lambda,m_{\omega},r),\\
\mathcal{V}_{I}^{\Xi_{cc}\Xi_c\to\Xi_{cc}\Xi_c^{\prime}} &=& -\frac{\mathcal{G}(I)}{12\sqrt{2}}\frac{g_{\pi}g_1}{f_{\pi}m_{\Xi_{cc}}}
\left[\mathcal{A}_1\mathcal{O}_r
+\mathcal{A}_2\mathcal{P}_r\right]Y(\Lambda_3,m_{\pi3},r)
-\frac{1}{12\sqrt{2}}\frac{g_{\pi}g_1}{f_{\pi}m_{\Xi_{cc}}}
\left[\mathcal{A}_1\mathcal{O}_r
+\mathcal{A}_2\mathcal{P}_r\right]Y(\Lambda_3,m_{\eta3},r)\nonumber\\
&&-\frac{\mathcal{G}(I)}{12\sqrt{3}}\frac{\lambda_Ig_V(h_V-f_V)}{m_{\Xi_{cc}}}
\left[2\mathcal{A}_1\mathcal{O}_r
-\mathcal{A}_2\mathcal{P}_r\right]Y(\Lambda_3,m_{\rho3},r)
+\frac{\mathcal{G}(I)}{2\sqrt{3}}\frac{\lambda_Ig_Vh_V}{m_{\Xi_{cc}}}\mathcal{A}_{3}\mathcal{Q}_r
Y(\Lambda_3,m_{\rho3},r)\nonumber\\
&&-\frac{1}{12\sqrt{3}}\frac{\lambda_Ig_V(h_V-f_V)}{m_{\Xi_{cc}}}\left[2\mathcal{A}_1\mathcal{O}_r
-\mathcal{A}_2\mathcal{P}_r\right]Y(\Lambda_3,m_{\omega3},r)
+\frac{1}{2\sqrt{3}}\frac{\lambda_Ig_Vh_V}{m_{\Xi_{cc}}}\mathcal{A}_{3}\mathcal{Q}_r
Y(\Lambda_3,m_{\omega3},r),\\
\mathcal{V}_{I}^{\Xi_{cc}\Xi_c\to\Xi_{cc}\Xi_c^*} &=& -\frac{\mathcal{G}(I)}{12\sqrt{2}}\frac{g_{\pi}g_4}{f_{\pi}m_{\Xi_{cc}}}
\left[\mathcal{A}_{8}\mathcal{O}_r
+\mathcal{A}_{9}\mathcal{P}_r\right]Y(\Lambda_4,m_{\pi4},r)
-\frac{1}{12\sqrt{2}}\frac{g_{\pi}g_4}{f_{\pi}m_{\Xi_{cc}}}
\left[\mathcal{A}_{8}\mathcal{O}_r
+\mathcal{A}_{9}\mathcal{P}_r\right]Y(\Lambda_4,m_{\eta4},r)\nonumber\\
&&+\frac{\mathcal{G}(I)}{4}\frac{\lambda_Ig_Vh_V}{m_{\Xi_{cc}}}
\mathcal{A}_{10}\mathcal{Q}_rY(\Lambda_4,m_{\rho4},r)
-\frac{\mathcal{G}(I)}{24}\frac{\lambda_Ig_V(h_V-f_V)}{m_{\Xi_{cc}}}
\left[2\mathcal{A}_{8}\mathcal{O}_r
-\mathcal{A}_{9}\mathcal{P}_r\right]Y(\Lambda_4,m_{\rho4},r)\nonumber\\
&&+\frac{1}{4}\frac{\lambda_Ig_Vh_V}{m_{\Xi_{cc}}}
\mathcal{A}_{10}\mathcal{Q}_rY(\Lambda_4,m_{\omega4},r)
-\frac{1}{24}\frac{\lambda_Ig_V(h_V-f_V)}{m_{\Xi_{cc}}}
\left[2\mathcal{A}_{8}\mathcal{O}_r
-\mathcal{A}_{9}\mathcal{P}_r\right]Y(\Lambda_4,m_{\omega4},r),\\
\mathcal{V}_{I}^{\Xi_{cc}\Xi_c^{\prime}\to\Xi_{cc}\Xi_c^{\prime}} &=& g_{\sigma}l_SY(\Lambda,m_{\sigma},r)
-\frac{\mathcal{G}(I)}{24}\frac{g_{\pi}g_1}{f_{\pi}M_{\Xi_{cc}}}\left(\mathcal{A}_1\mathcal{O}_r
+\mathcal{A}_2\mathcal{P}_r\right)Y(\Lambda,m_{\pi},r)
+\frac{g_{\pi}g_1}{72f_{\pi}M_{\Xi_{cc}}}\left(\mathcal{A}_1\mathcal{O}_r
+\mathcal{A}_2\mathcal{P}_r\right)Y(\Lambda,m_{\eta},r)\nonumber\\
&&-\frac{\mathcal{G}(I)}{4}\left(\frac{ h_v\beta_Sg_V}{\sqrt{2}}
+\frac{h_v\lambda_Sg_V}{3\sqrt{2}M_{\Sigma_c}}\mathcal{O}_r\right)Y(\Lambda,m_{\rho},r)
-\frac{\mathcal{G}(I)}{6\sqrt{2}}h_v\lambda_Sg_V
\left(\frac{1}{M_{\Sigma_c}}+\frac{1}{M_{\Xi_{cc}}}\right)
\mathcal{A}_3\mathcal{Q}_rY(\Lambda,m_{\rho},r)\nonumber\\
&&-\frac{\mathcal{G}(I)}{36\sqrt{2}}
\frac{(f_v+h_v)\lambda_Sg_V}{M_{\Xi_{cc}}}(2\mathcal{A}_1\mathcal{O}_r
-\mathcal{A}_2\mathcal{P}_r)Y(\Lambda,m_{\rho},r)\nonumber\\
&&-\frac{1}{4}\left(\frac{ h_v\beta_Sg_V}{\sqrt{2}}
+\frac{h_v\lambda_Sg_V}{3\sqrt{2}M_{\Sigma_c}}\mathcal{O}_r\right)Y(\Lambda,m_{\omega},r)
-\frac{1}{6\sqrt{2}}h_v\lambda_Sg_V
\left(\frac{1}{M_{\Sigma_c}}+\frac{1}{M_{\Xi_{cc}}}\right)
\mathcal{A}_3\mathcal{Q}_rY(\Lambda,m_{\omega},r)\nonumber\\
&&-\frac{1}{36\sqrt{2}}
\frac{(f_v+h_v)\lambda_Sg_V}{M_{\Xi_{cc}}}(2\mathcal{A}_1\mathcal{O}_r
-\mathcal{A}_2\mathcal{P}_r)Y(\Lambda,m_{\omega},r),\\
\mathcal{V}_{I}^{\Xi_{cc}\Xi_c^{\prime}\to\Xi_{cc}\Xi_c^*} &=& \frac{\mathcal{G}(I)}{16\sqrt{3}}\frac{g_{\pi}g_1}{f_{\pi}m_{\Xi_{cc}}}
\left[\mathcal{A}_{11}\mathcal{O}_r+\mathcal{A}_{12}\mathcal{P}_r\right] Y(\Lambda_5,m_{\pi5},r)
+\frac{1}{48\sqrt{3}}\frac{g_{\pi}g_1}{f_{\pi}m_{\Xi_{cc}}}
\left[\mathcal{A}_{11}\mathcal{O}_r+\mathcal{A}_{12}\mathcal{P}_r\right]
Y(\Lambda_5,m_{\eta5},r)\nonumber\\
&&+\frac{\mathcal{G}(I)}{4\sqrt{6}}\frac{\lambda_Sg_Vh_V}{m_{\Xi_{cc}}}\mathcal{A}_{13}
\mathcal{Q}_rY(\Lambda_5,m_{\rho5},r)
-\frac{\mathcal{G}(I)}{24\sqrt{6}}\frac{\lambda_Sg_V(f_V-h_V)}{m_{\Xi_{cc}}}
\left[2\mathcal{A}_{11}\mathcal{O}_r-\mathcal{A}_{12}\mathcal{P}_r\right]
Y(\Lambda_5,m_{\rho5},r)\nonumber\\
&&-\frac{1}{4\sqrt{6}}\frac{\lambda_Sg_Vh_V}{m_{\Xi_{cc}}}\mathcal{A}_{13}
\mathcal{Q}_rY(\Lambda_5,m_{\omega5},r)
+\frac{1}{24\sqrt{6}}\frac{\lambda_Sg_V(f_V-h_V)}{m_{\Xi_{cc}}}\left[2\mathcal{A}_{11}\mathcal{O}_r
-\mathcal{A}_{12}\mathcal{P}_r\right]Y(\Lambda_5,m_{\omega5},r),\\
\mathcal{V}_{I}^{\Xi_{cc}\Xi_c^{*}\to\Xi_{cc}\Xi_c^{*}} &=& g_{\sigma}l_S\mathcal{A}_4Y(\Lambda,m_{\sigma},r)
-\frac{\mathcal{G}(I)}{16}\frac{g_{\pi}g_1}{M_{\Xi_{cc}}f_{\pi}}
\left[\mathcal{A}_5\mathcal{O}_r+\mathcal{A}_6\mathcal{P}_r\right]Y(\Lambda,m_{\pi},r)
+\frac{1}{48}\frac{g_{\pi}g_1}{M_{\Xi_{cc}}f_{\pi}}
\left[\mathcal{A}_5\mathcal{O}_r+\mathcal{A}_6\mathcal{P}_r\right]Y(\Lambda,m_{\eta},r)\nonumber\\
&&-\frac{\mathcal{G}(I)}{4\sqrt{2}}h_v\beta_Sg_V
\mathcal{A}_4Y(\Lambda,m_{\rho},r)
-\frac{\mathcal{G}(I)}{4\sqrt{2}}\frac{h_v\lambda_Sg_V}{M_{\Xi_{cc}}}\mathcal{A}_7\mathcal{Q}_rY(\Lambda,m_{\rho},r)\nonumber\\
&&+\frac{\mathcal{G}(I)}{24\sqrt{2}}\frac{(h_v+f_v)\lambda_sg_V}{M_{\Xi_{cc}}}
\left[2\mathcal{A}_5\mathcal{O}_r-\mathcal{A}_6\mathcal{P}_r\right]Y(\Lambda,m_{\rho},r)\nonumber\\
&&-\frac{h_v\beta_Sg_V}{4\sqrt{2}}
\mathcal{A}_4Y(\Lambda,m_{\omega},r)
-\frac{h_v\lambda_Sg_V}{4\sqrt{2}}\mathcal{A}_7\mathcal{Q}_rY(\Lambda,m_{\omega},r)
+\frac{(h_v+f_v)\lambda_sg_V}{24\sqrt{2}M_{\Xi_{cc}}}
\left[2\mathcal{A}_5\mathcal{O}_r-\mathcal{A}_6\mathcal{P}_r\right]Y(\Lambda,m_{\omega},r)
\end{eqnarray}
\end{widetext}

with
\begin{eqnarray*}
&&\mathcal{O}_r = \frac{1}{r^2}\frac{\partial}{\partial r}r^2\frac{\partial}{\partial r},\quad
\mathcal{P}_r = r\frac{\partial}{\partial r}\frac{1}{r}\frac{\partial}{\partial r},\quad \mathcal{Q}_r=\frac{1}{r}\frac{\partial}{\partial r},\\
&&Y(\Lambda,m,r) = \frac{1}{4\pi r}(e^{-mr}-e^{-\Lambda r})-\frac{\Lambda^2-m^2}{8\pi \Lambda}e^{-\Lambda r}.
\end{eqnarray*}
The variables in above effective potentials denote
\begin{eqnarray*}\left.\begin{array}{lll}
q_0 = \frac{-m_{\Sigma_c}^2+m_{\Lambda_c}^2}{2(m_{\Xi_{cc}}+m_{\Sigma_c})},\quad\quad
&q_1= \frac{-m_{\Sigma_c^*}^2+m_{\Lambda_c}^2}{2(m_{\Xi_{cc}}+m_{\Sigma_c^*})},\quad\quad
&q_2= \frac{-m_{\Sigma_c^*}^2+m_{\Sigma_c}^2}{2(m_{\Xi_{cc}}+m_{\Sigma_c^*})},\\
q_3=\frac{-m_{\Xi^{\prime}_c}^2+m_{\Xi_c}^2}{2(m_{\Xi_{cc}}+m_{\Xi^{\prime}_c})},\quad\quad
&q_4= \frac{-m_{\Xi_c^*}^2+m_{\Xi_c}^2}{2(m_{\Xi_{cc}}+m_{\Xi_c^*})},\quad\quad
&q_5= \frac{-m_{\Xi_c^*}^2+m_{\Xi^{\prime}_c}^2}{2(m_{\Xi_{cc}}+m_{\Xi_c^*})},\\
\Lambda_i^2=\Lambda^2-q_i^2,
&m_{Ei}^2=m_{E}^2-q_i^2,
&i=0,1,2,3,4,5.\end{array}\right.
\end{eqnarray*}
Here, we define two isospin factors $\mathcal{G}(I)$ and $\mathcal{H}(I)$, respectively,
\begin{eqnarray}
\left.\begin{array}{ll}
{G}(I=0) =-3,\quad\quad\quad    &\mathcal{G}(I=1) =1,\\
\mathcal{H}(I=1/2) =-2,         &\mathcal{H}(I=3/2) =1.\end{array}\right.
\end{eqnarray}

In above potentials, we also define several spin-spin, spin-orbit, and tensor force operators, including
\begin{eqnarray}
\mathcal{A}_1 &=& \chi_4^{\dag}\chi_3^{\dag}\left(\bm{\sigma_1}\cdot\bm{\sigma_2}\right)\chi_2\chi_1,\\
\mathcal{A}_2 &=& \chi_4^{\dag}\chi_3^{\dag}S(\hat{r},\bm{\sigma_1},\bm{\sigma_2})\chi_2\chi_1,\\
\mathcal{A}_3 &=& \chi_4^{\dag}\chi_3^{\dag}\left(\bm{\sigma_{2}}\cdot\bm{L}\right)\chi_2\chi_1,\\
\mathcal{A}_4 &=&\sum_{a,b}^{m,n}C_{1/2,a;1,b}^{3/2,a+b}
  C_{{1}/{2},c;1,d}^{{3}/{2},c+d}\chi_{4a}^{\dag}\chi_3^{\dag}\chi_{2m}\chi_1
     \bm{\epsilon_2^n}\cdot\bm{\epsilon_4^{b\dag}},\\
\mathcal{A}_5 &=&\sum_{a,b}^{m,n}C_{{1}/{2},a;1,b}^{{3}/{2},a+b}
  C_{{1}/{2},c;1,d}^{{3}/{2},c+d}\chi_{4a}^{\dag}\chi_3^{\dag}
  i\bm{\sigma_1}\cdot\left(\bm{\epsilon_2^n}\times\bm{\epsilon_4^{b\dag}}\right)\chi_{2m}\chi_1,
  \nonumber\\\\
\mathcal{A}_6 &=&\sum_{a,b}^{m,n}C_{{1}/{2},a;1,b}^{{3}/{2},a+b}
  C_{{1}/{2},c;1d}^{{3}/{2},c+d}\chi_{4a}^{\dag}\chi_3^{\dag}
  S(\hat{r},\bm{\sigma_1},i\bm{\epsilon_2^n}\times\bm{\epsilon_4^{b\dag}})\chi_{2m}\chi_1,
  \nonumber\\\\
\mathcal{A}_7 &=&\sum_{a,b}^{m,n}C_{{1}/{2},a;1,b}^{{3}/{2},a+b}
  C_{{1}/{2},c;1,d}^{{3}/{2},c+d}\chi_{4a}^{\dag}\chi_3^{\dag}
  i\left(\bm{\epsilon_2^n}\times\bm{\epsilon_4^{b\dag}}\right)\cdot\bm{L}
  \chi_{2m}\chi_1,\nonumber\\\\
\mathcal{A}_{8} &=& \sum_{m,n}C_{1/2,m;1,n}^{3/2,m+n}\chi_{4m}^{\dag}\chi_3^{\dag}
\left(\bm{\sigma_1}\cdot\bm{\epsilon_4^{n\dag}}\right)\chi_2\chi_1,\\
\mathcal{A}_{9} &=& \sum_{m,n}C_{1/2,m;1,n}^{3/2,m+n}\chi_{4m}^{\dag}\chi_3^{\dag}
S\left(\hat{r},\bm{\sigma_1},\bm{\epsilon_4^{n\dag}}\right)\chi_2\chi_1,\\
\mathcal{A}_{10} &=& \sum_{m,n}C_{1/2,m;1,n}^{3/2,m+n}\chi_{4m}^{\dag}\chi_3^{\dag}
\left(\bm{L}\cdot\bm{\epsilon_4^{n\dag}}\right)\chi_2\chi_1,\\
\mathcal{A}_{11} &=& \sum_{m,n}C_{1/2,m;1,n}^{3/2,m+n}\chi_{4m}^{\dag}\chi_3^{\dag}
\left[\bm{\sigma_1}\cdot\left(i\bm{\sigma_2}\times\bm{\epsilon_4^{n\dag}}\right)\right]\chi_2\chi_1,\\
\mathcal{A}_{12} &=& \sum_{m,n}C_{1/2,m;1,n}^{3/2,m+n}\chi_{4m}^{\dag}\chi_3^{\dag}
S\left(\hat{r},\bm{\sigma_1},i\bm{\sigma_2}\times\bm{\epsilon_4^{n\dag}}\right)\chi_2\chi_1,\nonumber\\\\
\mathcal{A}_{13} &=& \sum_{m,n}C_{1/2,m;1,n}^{3/2,m+n}\chi_{4m}^{\dag}\chi_3^{\dag}
\left[\bm{L}\cdot\left(i\bm{\sigma_2}\times\bm{\epsilon_4^{n\dag}}\right)\right]\chi_2\chi_1,
\end{eqnarray}
with
\begin{eqnarray*}
S(\hat{r},\bm{a},\bm{b}) &=& 3(\hat{r}\cdot\bm{a})(\hat{r}\cdot\bm{b})-\bm{a}\cdot\bm{b}.
\end{eqnarray*}

In Table \ref{operator}, we present the corresponding matrices elements, which are obtained by sandwiched these operators between the relevant spin-orbit wave functions.
\renewcommand\tabcolsep{0.3cm}
\renewcommand{\arraystretch}{1.5}
\begin{table*}[!htbb]
\caption{Matrices elements for $\langle f|\mathcal{A}|i\rangle$ for the spin-spin, spin-orbit, and tensor force operators $\mathcal{A}$. For example, $\langle {}^1S_0|\mathcal{A}_1|{}^1S_0\rangle=-3$.}\label{operator}
{\begin{tabular}{clllll}
\toprule[1pt]
{Spin}
       &$\langle\mathcal{A}_1\rangle$       &$\langle\mathcal{A}_2\rangle$
       &$\langle\mathcal{A}_3\rangle$       &$\langle\mathcal{A}_4\rangle$
       &$\langle\mathcal{A}_5\rangle$       \\\hline
$J=0$   &$\left(-3\right)$
              &$\left(0\right)$
                    &$\left(0\right)$
                         &$\left(1\right)$
                             &$\left(1\right)$
                                 \\
$J=1$    &$\left(\begin{array}{cc}1  &0\\  0   &1 \end{array}\right)$
             &{$\left(\begin{array}{cc}0  &\sqrt{8}\\  \sqrt{8}   &-2 \end{array}\right)$}
                  &$\left(\begin{array}{cc}0  &0\\  0   &-{3} \end{array}\right)$
                      &$\left(\begin{array}{ccc} 1 & 0 & 0 \\ 0 & 1 & 0 \\ 0 & 0 & 1\end{array}\right)$
                          &{$\left(\begin{array}{ccc} -\frac{5}{3} & 0 & 0 \\ 0 & -\frac{5}{3} & 0 \\ 0 & 0 & 1\end{array}\right)$}
                                 \\
$J=2$    &$\left(\begin{array}{cc} -3 & 0 \\ 0 & 1 \\\end{array}\right)$
          &$\left(\begin{array}{cc} 0 & 0 \\ 0 & 2 \\\end{array}\right)$
            &{{$\left(\begin{array}{cc} 0 & -\sqrt{6} \\ -\sqrt{6} & -1 \\\end{array}\right)$}}
           &$\left(\begin{array}{ccc} 1 & 0 & 0 \\ 0 & 1 & 0 \\ 0 & 0 & 1\end{array}\right)$
               &$\left(\begin{array}{ccc} 1 & 0 & 0 \\ 0 & -\frac{5}{3} & 0 \\ 0 & 0 & 1\end{array}\right)$\\
\bottomrule[1pt]
{Spin}
       &$\langle\mathcal{A}_6\rangle$             &$\langle\mathcal{A}_7\rangle$
       &$\langle\mathcal{A}_{8}\rangle=-\langle\mathcal{A}_{11}\rangle$
       &$\langle\mathcal{A}_{9}\rangle=-\langle\mathcal{A}_{12}\rangle$
       &$\langle\mathcal{A}_{10}\rangle=-\langle\mathcal{A}_{13}\rangle$\\\hline
$J=0$   &$\left(-2\right)$
               &$\left(3\right)$
                   &$\left(0\right)$
           &$\langle\mathcal{A}_{9}\rangle=\left(\sqrt{6}\right)$, $\langle\mathcal{A}_{12}\rangle=(0)$
              &$\left(0\right)$\\
$J=1$    &{$\left(\begin{array}{ccc} 0 & -\frac{\sqrt{2}}{3} & -\sqrt{2} \\ -\frac{\sqrt{2}}{3} & \frac{1}{3} & -1 \\-\sqrt{2} & -1 & -1\end{array}\right)$}
                                    &$\left(\begin{array}{ccc} 0 & 0 & 0 \\0 & \frac{5}{2} & -\frac{1}{2} \\
           0 & -\frac{1}{2} & \frac{5}{2}\end{array}\right)$
                &$\left(\begin{array}{cc} 2 \sqrt{\frac{2}{3}} & 0 \\ 0 & 2 \sqrt{\frac{2}{3}} \\ 0 & 0 \\\end{array}\right)$
                        &$\left(\begin{array}{cc} 0 & -\frac{1}{\sqrt{3}} \\ -\frac{1}{\sqrt{3}} & \frac{1}{\sqrt{6}} \\ -\sqrt{3} & -\sqrt{\frac{3}{2}} \\\end{array}\right)$
                             &$\left(\begin{array}{cc} 0 & 0 \\ 0 & -\sqrt{\frac{3}{2}} \\ 0 & -\sqrt{\frac{3}{2}} \\\end{array}\right)$\\
$J=2$      &{$\left(\begin{array}{ccc} 0 & \sqrt{\frac{6}{5}} & \sqrt{\frac{14}{5}} \\ \sqrt{\frac{6}{5}} & -\frac{1}{3} & -\sqrt{\frac{3}{7}} \\ \sqrt{\frac{14}{5}} & -\sqrt{\frac{3}{7}} & \frac{3}{7}\end{array}\right)$}
                       &$\left(\begin{array}{ccc} 0 & 0 & 0 \\ 0 & \frac{5}{6} & -\sqrt{\frac{7}{12}} \\ 0 & -\sqrt{\frac{7}{12}} & \frac{3}{2}\end{array}\right)$
                &$\left(\begin{array}{cc}0 & 0 \\ 0 & 2 \sqrt{\frac{2}{3}} \\ 0 & 0 \\\end{array}\right)$
                    &$\left(\begin{array}{cc} \sqrt{\frac{6}{5}} & \frac{3}{\sqrt{5}} \\ 0 & -\frac{1}{\sqrt{6}} \\ -\sqrt{\frac{3}{28}} & -\frac{3}{\sqrt{14}} \\\end{array}\right)$
                         &$\left(
\begin{array}{cc}
 0 & 0 \\
 -2 & -\frac{1}{\sqrt{6}} \\
 0 & -\sqrt{\frac{7}{2}} \\
\end{array}
\right)$
\\\bottomrule[1pt]
\end{tabular}}
\end{table*}

\end{document}